\documentclass[12pt]{article}

\pdfoutput=1

\usepackage[top=80pt,bottom=85pt,left=85pt,right=85pt]{geometry}
\usepackage{amssymb}
\usepackage{amsmath}
\usepackage{setspace}
\usepackage{sectsty}
\usepackage{cite}
\usepackage{bbm}
\usepackage{bm}
\usepackage[utf8]{inputenc}
\usepackage[T1]{fontenc}
\usepackage{comment}
\usepackage[english]{babel}
\usepackage{afterpage}
\usepackage{bbold}

\usepackage[usenames]{xcolor}
\usepackage{graphicx,subcaption}
\usepackage{setspace}
\usepackage{float}
\usepackage{booktabs}
\usepackage[debug,pageanchor=false]{hyperref}
\definecolor{link}{rgb}{.8,.15,.1}
\definecolor{pigment}{rgb}{0.36, 0.54, 0.66}
\definecolor{pigment2}{rgb}{0.19, 0.55, 0.91}
\definecolor{pigment3}{rgb}{0.2, 0.2, 0.6}
\definecolor{light-gray}{gray}{0.75}
\hypersetup{colorlinks=true,linkcolor=link,citecolor=link,urlcolor=link,linktocpage}

\usepackage{array,multirow,booktabs,longtable}
\usepackage{mathrsfs}
\usepackage{tikz-cd} 
\usetikzlibrary{backgrounds, arrows,calc,shapes,decorations.pathreplacing, decorations.markings, automata,positioning}
\tikzset{%
  >={Latex[width=2mm,length=2mm]},
            base/.style = {rectangle, rounded corners, draw=black,
                           minimum width=4cm, minimum heigwht=1cm,
                           text centered, font=\sffamily},
  activityStarts/.style = {base, fill=orange!15},
       startstop/.style = {base, fill=orange!15},
    activityRuns/.style = {base, fill=orange!15},
         process/.style = {base, minimum width=2.5cm, fill=orange!15,
                           font=\ttfamily},
}

\newcommand{\red}[1]{}
\renewcommand{\red}[1]{{\color{red} {#1}}}
\newcommand{\blue}[1]{{\color{blue} {#1}}}

\tikzset{
        cvertex/.style={circle,draw=black,inner sep=1pt,outer sep=3pt},
        vertex/.style={circle,fill=black,inner sep=1pt,outer sep=3pt},
        star/.style={circle,fill=yellow,inner sep=0.75pt,outer sep=0.75pt},
        tvertex/.style={inner sep=1pt,font=\scriptsize},
        gap/.style={inner sep=0.5pt,fill=white}}

\tikzstyle{mybox} = [draw=black, fill=blue!10, very thick,
    rectangle, rounded corners, inner sep=10pt, inner ysep=20pt]
\tikzstyle{boxtitle} =[fill=blue!50, text=white,rectangle,rounded corners]

\newcommand\scalemath[2]{\scalebox{#1}{\mbox{\ensuremath{\displaystyle #2}}}}

\newcommand{\todo}[1]{}
\renewcommand{\todo}[1]{{\color{red} TODO: {#1}}}
\renewcommand{\red}[1]{{\color{red} {#1}}}

\newcommand{\be}{\begin{equation}}  
\newcommand{\ee}{\end{equation}}  
\newcommand{\bea}{\begin{align}}
\newcommand{\eea}{\end{align}}
\newcommand{\bp}{\begin{bmatrix*}[r]}  
\newcommand{\ep}{\end{bmatrix*}}  
\newcommand{\bpp}{\begin{bmatrix}}  
\newcommand{\epp}{\end{bmatrix}}  
\newcommand{\bcd}{\begin{center}
\begin{tikzcd}}
\newcommand{\ecd}{\end{tikzcd} \end{center}}
\newcommand{\bbm}{\begin{pmatrix}}  
\newcommand{\eem}{\end{pmatrix}}

\makeatletter
\@addtoreset{equation}{section}
\makeatother

\begin{document}

\begin{titlepage}

\begin{center}

\vskip .3in \noindent

{\Large \bf{Flops of any {\it length}, Gopakumar-Vafa invariants and 5d Higgs Branches}}

\bigskip\bigskip

Andr\'es Collinucci$^a$, Mario De Marco$^{b}$, Andrea Sangiovanni$^{c}$ and Roberto Valandro$^{c}$ \\

\bigskip


\bigskip
{\footnotesize
 \it

$^a$ Service de Physique Th\'eorique et Math\'ematique, Universit\'e Libre de Bruxelles and \\ International Solvay Institutes, Campus Plaine C.P.~231, B-1050 Bruxelles, Belgium\\
\vspace{.25cm}
$^b$ SISSA and INFN, Via Bonomea 265, I-34136 Trieste, Italy\\
\vspace{.25cm}
$^c$ Dipartimento di Fisica, Universit\`a di Trieste, Strada Costiera 11, I-34151 Trieste, Italy \\
and INFN, Sezione di Trieste, Via Valerio 2, I-34127 Trieste, Italy	
}

\vskip .5cm
{\scriptsize \tt collinucci dot phys at gmail dot com \hspace{2cm}  mdemarco at sissa dot it} \\
{\scriptsize \tt \hspace{0.5cm}  andrea dot sangiovanni at phd dot units dot it \hspace{1cm}  roberto dot valandro at ts dot infn dot it}

\vskip 1cm
     	{\bf Abstract }
	\end{center}
\vskip .1in
The conifold is a basic example of a noncompact Calabi-Yau threefold that admits a simple flop, and in M-theory, gives rise to a 5d hypermultiplet at low energies, realized by an M2-brane wrapped on the vanishing sphere. 
We develop a novel gauge-theoretic method to construct
new classes of examples that generalize the simple flop to so-called \emph{length} $\ell=1, \ldots, 6$. 
The method allows us to naturally read off the Gopakumar-Vafa invariants. Although they share similar properties to the beloved conifold, these threefolds are expected to admit M2-bound states of higher degree $\ell$. We demonstrate this through our computations of the GV invariants. Furthermore we fully characterize the associated Higgs branches by computing their dimensions and flavor groups.
With our techniques we extract more refined data such as the charges of the hypers under the flavor group.

\noindent

\vfill
\eject

\end{titlepage}

\tableofcontents

\section{Introduction}\label{Sec:Intro}

Singular noncompact Calabi-Yau threefolds are interesting objects both from a mathematical as well as a physics point of view. In physics, by defining string theory or M-theory on them, they can provide the framework for modeling effective supersymmetric gauge theories in four and five dimensions, with gravity decoupled. This can happen via the \emph{branes at singularities} paradigm, whereby a bunch of D3-branes probes such a singularity, or via the \emph{geometric engineering} paradigm, whereby the field theory content emanates from membranes wrapping the vanishing cycles of the singularity. We will focus on the latter.

The prototypical example, the \emph{conifold}, is a threefold that can be described as a hypersurface in $\mathbb{C}^4$. It admits a so-called \emph{small resolution}, which means that the singularity is replaced by a $\mathbb{P}^1$ (unlike other singularities where the exceptional locus is of complex dimension two). This sphere can flop to negative volume, and a homologically inequivalent sphere of positive area grows.

The conifold is but the first in a series of threefolds that admit \emph{simple flops}. Here, `simple' refers to the fact that only one sphere grows after resolution (as opposed to a chain of spheres). The conifold is known as a simple flop of \emph{length one}. Here, the \emph{length} of a flop refers to a multiplicity that can be attributed to the exceptional $\mathbb{P}^1$. If one pulls back the skyscraper sheaf corresponding to the singular point of the conifold w.r.t. the blow-down map, one will get the structure sheaf of the exceptional curve. 
\begin{equation}
\pi^*(\mathcal{O}_p) = \mathcal{O}_{\mathbb{P}^1}\,.
\end{equation}
It has been known for decades that other \emph{lengths} are possible, up to $l= 6$, such that
\begin{equation}
\pi^*(\mathcal{O}_p) = \mathcal{O}_{\mathbb{P}^1}^{\oplus l}\,.
\end{equation}
Even though the above definition may appear rather obscure, the length has a very concrete physical meaning: as we will see later, when we geometrically engineer M-theory on simple threefold flops described as deformed ADE singularities, we find that the length of the simple flop is nothing but the \textit{maximal flavor charge} of the 5d hypermultiplets arising from M2 branes wrapped on the exceptional $\mathbb{P}^1$. We will get into the details of these aspects in later sections.

In order to construct such higher length flops, the strategy in \cite{Katz:1992aa} has been to consider families of deformed ADE surfaces. In other words, fibering local K3's over the deformation parameter space. In general, the total space of the families may or may not be singular. Resolving the possible singularity inflates part of the exceptional $\mathbb{P}^1$'s associated to the simple roots of the ADE algebra associated with the family. Given such a family, which will give rise to an $n$-fold, one can always cut out a complex three-dimensional subvariety, which will be guaranteed to give a Gorenstein threefold.

\

In this paper, following \cite{Collinucci:2021wty,Collinucci:2021ofd}, we develop a new method to construct simple flops spanning all the possible values of the length of the exceptional $\mathbb{P}^1$ (up to the maximal value $l=6$), based on a gauge-theoretic approach.
To this end, we consider M-theory on an ALE surface with an ADE singularity. The 7d effective theory is a SYM theory with three adjoint scalars and group $G=A,D,E$ with $\mathcal{N}=1$ supersymmetry. One can break half of the supercharges by switching on a BPS vev for the scalar fields in the following way:
\begin{itemize}
\item Take the 7d spacetime to be $\mathbb{R}^5\times \mathbb{C}$. Call $w$ the local coordinate on $\mathbb{C}$.
\item Split the tree real scalars into a complex adjoint scalar $\Phi=\phi_1+i\phi_2$ and a real adjoint scalar $\phi_3$.
\item Take a holomorphic vev for $\Phi$ that depends on a coordinate $w$, transverse to the local K3.
\end{itemize}
This preserves only a 5d Poincar\'e group. The 7d gauge group is broken to the commutant of $\Phi$. The 7d $\mathcal{N}=1$  vector multiplet was made up of the gauge field $A_M$ ($M=0,1,...,6$) and the three scalars $\phi_i$ ($i=1,2,3$); after giving a vev to $\Phi$, only a 5d $\mathcal{N}=1$  vector multiplet $(A_\mu,\phi_3)$ in the subalgebra $\mathcal{H}$ that commutes with $\Phi$ survives (these vector multiplets still propagate in 7d, and are then seen as background vector multiplets from the 5d point of view).
As we will see in detail in the main text, there are also localized $\Phi$-modes at $w=0$ that organize in hypermultiplets charged under $\mathcal{H}$ and that propagate in 5d. 

The fields $\Phi$ and $\phi_3$ can be given two physical interpretations: In the A or D series, the local K3 has a $\mathbb{C}^*$-fibration structure; we can then reduce M-theory to IIA, and regard them as the three adjoint scalars or a stack of D6-branes. For the E series, we can consider F-theory on an elliptically fibered K3, giving rise to 8d SYM. Physically, $\Phi$ is the adjoint Higgs on the stack of non-perturbative sevenbranes. By further compactifying on a circle, we arrive at 7d SYM with exceptional gauge group.

The vev's of $\Phi$ and $\phi$ encode the complex structure and K\"ahler moduli of the local K3, respectively. When both are zero, the surface is singular. Activating $\phi$ blows up spheres of non-zero K\"ahler volume. However, activating $\Phi$ will switch on \emph{versal deformations} of the local K3, which will unfold the singularity according to its Casimir invariants.
From the geometric point of view, we will obtain a fibration of deformed ALE surfaces over the $w$-plane, giving rise to a threefold $X$ that admits a simple flop. Reducing M-theory directly on $X$ produces the 5d spectrum of propagating hypermultiplets and background vector multiplets outlined before.

Resolving the full fibration is known as a \emph{simultaneous resolution}.
Given the Dynkin diagram of the gauge algebra in question, a simultaneous resolution corresponds to a choice of simple roots. This selects, the subspace $\mathcal{H}$ of the Cartan subalgebra $\mathfrak{t}$ of the original ADE algebra where $\phi_3$ lives. 
The commutant of $\phi_3$ tells us what is the form of a generic $\langle\Phi\rangle$ producing the sought-after family. The M-theory threefold is then obtained by choosing the $w$-dependence of the Casimir invariants of $\Phi$, which uniquely fix the defining equation of the threefold. 

At this point, one can compute the number of hypermultiplets of the 5d effective theory, by counting the zero modes of the complex adjoint $\Phi$ that are localized around the singularity. Their number corresponds to the BPS index of M2-states wrapping the vanishing $\mathbb{P}^1$'s, aka the Gopakumar-Vafa invariants. 
In this paper, we are able to compute the charge of each hypermultiplet with respect to the flavor group, which gives us not only the multiplicity of states, but the actual refined information of the GV invariants, which are organized by homology classes of the threefold. This furthermore allows us to compute detailed information about the 5d Higgs branch: Its dimension is equal to the sum of all GV invariants while the action of the flavor group is determined by the charges of the hypers.

The paper is organized as follows. In Section~\ref{Sec:ADEfamilies} we review the families of deformed ADE singularities. In Section~\ref{Sec:3folds} we explain how the choice of spheres that are blown up in the simultaneous resolution determines the vev for $\Phi$; by using this $\Phi$ we then show how to extract the hypersurface equation of the threefold and its GV invariants.
In Section~\ref{Simple flop section} we apply our method to simple flops, by constructing explicitly threefolds with any length from $1$ to $6$.
In Section~\ref{Sec:conclusions} we draw our conclusions.
We add four appendices: in Appendix~\ref{App:Threefold} we outline the Katz-Morrison method to derive the threefold equation; in Appendix~\ref{Appendix A} we write some long formulae for the exceptional cases, necessary to compute the corresponding threefold equations; in Appendix~\ref{App Milnor} we illustrate a check of our computations by using the Milnor number of the considered singularities;
in Appendix~\ref{Appendix B} we expose the Mathematica code we worked out to compute the zero modes once one chooses a specific $\Phi$.

\section{ADE families and simultaneous resolution}\label{Sec:ADEfamilies}
In this section, we briefly summarize the philosophy and method of \cite{Katz:1992aa} for building simultaneous resolutions, by fibering ALE surfaces over the Cartan torus of an ADE group.
Consider an ALE surface in $\mathbb{C}^3$ with an ADE singularity, defined by the singular hypersurface equation $f(x,y,z)=0$. The cases in the ADE classification can be presented as:
\begin{equation}\label{ADE singularities}
\begin{split}
& A_r: \hspace{1cm} x^2+y^2+z^{r+1}=0\\
& D_r: \hspace{1cm} x^2+z y^2+z^{r-1}=0\\
& E_6: \hspace{1cm} x^2+y^3+z^4=0 \\
& E_7: \hspace{1cm} x^2+y^3+yz^3=0 \\
& E_8: \hspace{1cm} x^2+y^3+z^5=0 \\
\end{split}
\end{equation}
One can deform these equations by adding a certain number of monomials:
\begin{equation}
f+\sum_{i=1}^k \mu_ig_i = 0,
\end{equation}
where the monomials $g_i$ belong to the ring:
\begin{equation}
R = \frac{\mathbb{C}(x,y,z)}{\left(\frac{\partial f}{\partial x},\frac{\partial f}{\partial y},\frac{\partial f}{\partial z}\right)}.
\end{equation}
The coefficients $\mu_i$ of the monomials $g_i$ are coordinates on the space of deformations. One can construct a space by fibering the deformed ALE space over the deformation space: at the origin (where all deformation coefficients are zero), the ALE space develops the ADE singularity. The total space of the fibration may or may not be singular, depending on the actual fibration structure.

To illustrate the setup in more detail, we focus on an $A_{r-1}$ type ALE family. This machinery is well-developed in \cite{Katz:1992aa}. The equation defining the $A_{r-1}$ singularity is\footnote{Obtained from \eqref{ADE singularities} by the change of variables $x=\frac{v-u}{2}, y=\frac{v+u}{2i}$.}
\be 
 uv=z^r \:.
\ee
The versal deformation is
\be \label{Anfamily}
 uv=z^r + \sum _{i=2}^r (-1)^{i-1}\sigma_i z^{n-i}\:.
\ee
Promoting the $\sigma_i$'s to coordinates, this equation describes a fibration with fiber given by the (deformed) $A_{r-1}$ surface and  base the space $\mathfrak{t}/\mathcal{W}$ of gauge invariant coordinates $\sigma_i$ ($i=2,...,r$)\footnote{For $A_{r-1}$, the invariant $\sigma_i$ is the $i$-th elementary symmetric polynomial in the eigenvalues of an element of the Lie algebra.}
on the Lie algebra $\mathfrak{sl}(r)$, where $\mathfrak{t}$ is the  Cartan torus and $\mathcal{W}$ the Weyl group. 

The space \eqref{Anfamily}, a fibration over $\mathfrak{t}/\mathcal{W}$, is non-singular. However, by making a \emph{base change}, one obtains a singular space whose resolution blows up a subset of the roots of the central ALE fiber. Let us consider the simplest case, i.e. $r=2$. The equation \eqref{Anfamily} becomes
 \be 
 uv=z^2 + \sigma_2 \:,
\ee
which is perfectly smooth as a total space, even though the central fiber over $\sigma_2=0$ is a singular $A_1$-surface.
If we make the base change $\sigma_2=t^2$, i.e. we pull-back the ALE fibration w.r.t. the map:
\begin{align}
\mathfrak{t} &\longrightarrow \mathfrak{t}/\mathbb{Z}_2\\
t & \mapsto \sigma_2=t^2\,,
\end{align}
where, in this case, $\mathfrak{t} \cong \mathbb{C}$, we obtain a threefold with a conifold singularity:
\begin{equation}\label{conifold}
uv=z^2 + t^2\:.
\end{equation} 
The small resolution of the conifold blows up the simple root of $A_1$ in the central fiber. This is called a \emph{simultaneous resolution}.
The family is now fibered over the base $\mathfrak{t}$, whose $t$ is a coordinate. 

For generic $r$ one can make an analogous base change by pulling back the family w.r.t. to the map
\begin{equation}
\mathfrak{t} \longrightarrow \mathfrak{t}/\mathcal{W} \qquad t_i \mapsto \sigma_k(t_i)
\end{equation}
by taking $\sigma_k$'s written as $\mathcal{W}$-invariant functions of coordinates $t_i$ on  $\mathfrak{t}$. This resolves all the simple roots of $A_{r-1}$ in the central fiber.
One can also choose a different base change where
\begin{align}
\mathfrak{t} &&\longrightarrow& &\mathfrak{t}/\mathcal{W}' &&\longrightarrow &&\mathfrak{t}/\mathcal{W}&\\
t_i && \mapsto &&\varrho_i(t_j) &&\mapsto &&\sigma_i(t_j)&
\end{align}
where $\mathcal{W}'\subset \mathcal{W}$. In this case, the resolution of the family blows up the roots that are left invariant by $\mathcal{W}'$, in the central fiber. The base of the fibration is now parametrized by the $r-1$ $\mathcal{W}'$ invariants, that we call~$\varrho_i$~($i=1,...,r-1$). The $\varrho_i$'s can also be associated with $\mathcal{W}'$-invariant functions of coordinates $t_i$ on $\mathfrak{t}$.

This line of reasoning can be generalized \cite{Katz:1992aa} to all the cases in the ADE classification. 
Let us consider the family fibered over $\mathfrak{t}$.
At the origin of $\mathfrak{t}$, the surface develops the corresponding ADE singularity. At a generic point of $\mathfrak{t}$, the singularity is deformed: the surface admits $n$ non-vanishing $S^2$ intersecting in the same pattern as the nodes of the Dynkin diagram and whose volume is measured by the holomorphic (2,0)-form. 
In the deformed ALE surface, the holomorphic (2,0)-form is along an element ${\bf t}$  of the Cartan torus, such that ${\rm vol}_{\alpha_i}=\int_{\alpha_i}\Omega_{2,0}= \alpha_i[{\bf t}]$.
We choose a set of coordinates $t_i$ of $\mathfrak{t}$ such that the volumes of the simple roots are given by
\begin{equation}\label{volumes}
\begin{split}
& A_{r}:\quad {\rm vol}_{\alpha_i} = t_i-t_{i+1} \qquad i=1,...,r
\\
& D_{r}:\quad {\rm vol}_{\alpha_i} = \left\{\begin{array}{l}
 t_i-t_{i+1} \qquad i=1,...,r-1 \\  t_{r-1}+t_{r} \qquad i=r 
\end{array}\right.
 \\
& E_{r}: \quad  {\rm vol}_{\alpha_i} = \left\{\begin{array}{l}
 t_i-t_{i+1} \qquad i=1,...,r-1 \\  -t_{1}-t_{2}-t_{3} \qquad i=r
\end{array}\right.
\end{split}
\end{equation}
It is useful to state the explicit form of the versal deformations of the singularities \eqref{ADE singularities} as fibrations of deformed ADE surfaces over the base $\mathfrak{t}$ (with simultaneous resolution of all the simple roots):
\begin{equation}\label{deformed ADE singularities}
\begin{split}
& A_{r}:\quad x^{2}+y^{2}+\prod_{i=1}^{r+1}\left(z-t_{i}\right)=0 \quad\quad \sum_{i=1}^{r+1} t_{i}=0 \\
& D_{r}:\quad x^{2}+ zy^{2}+\frac{\prod_{i=1}^{r}\left(z+t_{i}^{2}\right)-\prod_{i=1}^{r} t_{i}^{2}}{z}+2 \prod_{i=1}^{r} t_{i} y=0 \\
& E_{6}: \quad x^{2}+z^{4}+y^{3}+\epsilon_{2} y z^{2}+\epsilon_{5} y z+\epsilon_{6} z^{2}+\epsilon_{8} y+\epsilon_{9} z+\epsilon_{12} =0\\
& E_{7}: \quad x^{2}+y^{3}+y z^{3}+\tilde{\epsilon}_{2} y^{2} z+\tilde{\epsilon}_{6} y^{2}+\tilde{\epsilon}_{8} y z+\tilde{\epsilon}_{10} z^{2}+\tilde{\epsilon}_{12} y+\tilde{\epsilon}_{14} z+\tilde{\epsilon}_{18}=0 \\
& E_{8}: \quad x^{2}+y^{3}+z^{5}+\hat{\epsilon}_{2} y z^{3}+\hat{\epsilon}_{8} y z^{2}+\hat{\epsilon}_{12} z^{3}+\hat{\epsilon}_{14} y z+\hat{\epsilon}_{18} z^{2}+\hat{\epsilon}_{20} y+\hat{\epsilon}_{24} z+\hat{\epsilon}_{30}=0,
\end{split}
\end{equation}
where the $\epsilon_i,\tilde{\epsilon}_i,\hat{\epsilon}_i$ are known functions of the parameters $t_i \in \mathfrak{t}$ (see \cite{Katz:1992aa} for the explicit expression of $\epsilon_i$, $\tilde{\epsilon}_i$ and an algorithm to compute $\hat{\epsilon}_i$).

The forms \eqref{deformed ADE singularities} admit the blow up of a collection of $\mathbb{P}^1$s in the central fiber dual to \textit{all} the simple roots of the corresponding ADE algebra. Exactly as for the $A_r$-families, one can make a simultaneous resolution of a subset of the simple roots, by choosing a subgroup $\mathcal{W}'\subset\mathcal{W}$ that leaves these roots invariant. The corresponding family is fibered over the space parametrized by the coordinates $\varrho_i$ of $ \mathfrak{t}/\mathcal{W}'$.

\section{Threefolds as ALE families, GV invariants and $\Phi$}\label{Sec:3folds}

The families of ALE surfaces described in the previous section are a natural starting point for constructing CY three-folds:  the coordinates $\varrho_i \in \mathfrak{t}/\mathcal{W}'$ can be chosen so as to depend (linearly)\footnote{Other dependences are indeed possible. However, we will stick to linear in order to avoid creating further singularities in the resulting threefolds.} on a single complex parameter $w$.
Such a threefold has then the structure of an ALE fibration over the complex plane $\mathbb{C}_w$ and it will be defined by an equation such as
\begin{equation}
 F\left(x,y,z,\varrho_1(w),...,\varrho_r(w)\right) = 0 \:,
\end{equation}
where $F\left(x,y,z,\varrho_1,...,\varrho_r\right) = 0 $ is the defining equation of the $(r+2)$-dimensional family ($r$ is the rank of the ADE algebra).
We will restrict on threefolds $X$ that have isolated singularities,
whose exceptional locus does not contain compact divisors.\footnote{One would need a  severe degeneration of the ALE surface at the origin, e.g.\ the ALE surface should split into several components, while in our cases the ALE surface just develops a singularity.}

In the following we are going to show how to build a  threefold family of deformed ADE surfaces by starting from the requirement that the  ALE fibration over $\mathbb{C}_w$ presents a specific partial simultaneous resolution, i.e.\ a choice of resolved simple roots, say $\alpha_1,...\alpha_\ell$. 
As we will see, this method can actually produce in principle the full ADE family over $\mathfrak{t}/\mathcal{W}'$.

\subsection{The Higgs field from the simultaneous resolution}
\label{Sec:ConstructionMethod}

The ADE families over $\mathfrak{t}/\mathcal{W}'$ and the corresponding threefolds have been derived and studied using several methods, such as \cite{Katz:1992aa,Curto:aa,Karmazyn:2017aa,Cachazo:2001gh,Collinucci:2018aho,Collinucci:2019fnh}. 
We are going to use a different (more physical) approach to construct these spaces that is based on the choice of an element $\Phi$ of the ADE algebra, depending on the coordinate $w$. 
Our starting point is M-theory reduced on an ALE surface with an ADE singularity of type $\mathfrak{g}$. 
M-theory on an ADE singularity gives rise to a $\mathcal{N}=1$, $d=7$ gauge theory with gauge group $G_{ADE}$ and three adjoint scalars $\phi_i$ ($i=1,2,3$).  
With a choice of a complex structure, a vev for $\Phi\equiv \phi_1+i\phi_2$ corresponds geometrically to a deformation of the singularity, while a vev for $\phi_3$ corresponds to a resolution, with $[\langle \Phi \rangle,\langle\phi_3\rangle]=0$. In particular the coefficients of the deforming monomials are associated with the Casimir invariants of $\Phi$.

We fiber this background over the $\mathbb{C}_w$-plane by switching on a non-zero $w$-dependent vev for the complex adjoint scalar $\Phi$. Geometrically this corresponds to deforming the singularity differently at different values of $w$, obtaining a threefold $X$ that is an ALE fibration over $\mathbb{C}_w$.

We now want to select a vev for $\Phi$ that allows the (simultaneous) resolution at the origin only of a choice of simple roots  $\alpha_1,...\alpha_\ell$ of $\mathfrak{g}$. Rephrased in the 7d field theory language,  $\langle\Phi\rangle$ must  be compatible with switching on a vev for $\phi_3$ along the subalgebra
\begin{equation}\label{calH}
\mathcal{H}=\langle\alpha_1^*,...\alpha_\ell^*\rangle\:.
\end{equation}
Since  $[\langle \Phi \rangle,\langle\phi_3\rangle]=0$, then $\Phi$ must live in the commutant of $\mathcal{H}$,  that we call $\mathcal{L}$. 
%
Such a subalgebra, that is the commutant of diagonal elements, is called a Levi subalgebra~of~ $\mathfrak{g}$.\footnote{At the level of the effective 7d theory, the vev $\langle\Phi\rangle$ breaks the gauge algebra to its commutant $\mathcal{H}$. It is important that  $\mathcal{H}$ is a subspace of the Cartan subalgebra of $\mathfrak{g}$.
In fact, if the preserved algebra contained  a simple factor $\mathfrak{g}'$, the vev $\langle\Phi\rangle$ would not deform the ADE singularity completely, but the fiber over generic $w$ would have a singularity of type $\mathfrak{g}'$, i.e. $X$ would have a non-isolated singularities.}  

%

Summing up, the choice of the blown up simple roots select an abelian subalgebra $\mathcal{H}\subset \mathfrak{g}$. This defines a Levi subalgebra $\mathcal{L}\subset\mathfrak{g}$, that is of the form
\begin{equation}\label{PhiInL}
\mathcal{L}= \bigoplus_{h} \mathcal{L}_h \oplus \mathcal{H} 
\end{equation}
with $\mathcal{L}_h$ simple Lie algebras.
The Higgs field\footnote{For ease of notation, from now on we call $\Phi$ the vev of the field corresponding to a given family of ADE deformed singularities.} $\Phi$ corresponding to the ALE fibration should be a generic element of $\mathcal{L}$.

The Casimir invariants of the Higgs field $\Phi$ tell us how the ALE fiber is deformed. Since at the origin of $\mathbb{C}_w$ the fiber presents the full ADE singularity, we expect that all the Casimir invariants of $\Phi$ vanish at $w=0$.
This means that at $w=0$ the Higgs field should be a non-zero nilpotent element of $\mathcal{L}$. 
In particular, when restricted on each summand $\mathcal{L}_h$ of the Levi subalgebra, $\Phi$ must be in the corresponding principal nilpotent orbit,  if we do not want terminal singularities.\footnote{If this does not happen, there will be neutral zero modes of $\Phi$, that, as we will see, will correspond to hypermultiplets that cannot get mass by going to the Coulomb branch; this corresponds to non-resolvable singularities.} 


We call the Higgs at the origin $X_+$. %
The generic Higgs field has then the following form 
\begin{equation}\label{PhiXY}
\Phi = X_+ + w\,Y
\end{equation}
where $Y\in\mathcal{L}$ can in principle depend on $w$; in this paper we mainly consider cases with constant $Y$.

The Higgs field $\Phi$ must deform the singularity outside the origin. 
At least in a neighborhood of $w=0$, we demand that the ALE fiber does not develop any singularity. This happens when $\Phi$ restricted on each summand of $\mathcal{L}$ is a non-zero semisimple element of that summand.
For $\mathcal{L}_h=A_{m-1}$, the generic form of $\Phi$, up to gauge transformations is in the form of a \emph{reconstructible Higgs}  \cite{Cecotti:2010bp}.
\be\label{recHiggsU}
\Phi|_{A_{ m-1}} =\left(\begin{array}{ccccc}
0 & 1 & 0 & \cdots & 0 \\
0 & 0 & 1 & 0 & 0\\
\vdots & 0 & \ddots & \ddots & 0  \\
0 & 0 & 0 & 0 & 1 \\
(-1)^{m-1}\hat\sigma_m & (-1)^{m-2}\hat\sigma_{m-1} & \cdots & -\hat\sigma_2& 0  \\
\end{array} \right)
\ee
with $\hat\sigma_j$ ($j=2,...,m$) the Casimirs of $\Phi|_{A_{ m-1}} $ (we called them the partial Casimir of $\mathfrak{g}$). 
There are analogous canonical forms when the summand is a different Lie algebra. 
Collecting the Casimirs  $\hat\sigma_j$'s for each summand $\mathcal{L}_h$ and the coefficient deformations along $\mathcal{H}$ one obtains the set of invariant coordinates  $\varrho_i$ that span the base of the family with simultaneous partial resolution.
The total Casimirs of $\Phi$ (that appear as coefficients of the deforming monomials in the versal deformation of the ADE singularity) can be written as functions of the $\varrho_i$'s.
Notice that by formulae like \eqref{recHiggsU}, we give $\Phi$ as a function of the partial Casimirs $\varrho_i$. The  choice of a dependence of $\varrho_i$ on $w$ produces a threefold.

We finally notice that the ADE Lie Algebra $\mathfrak{g}$ can be decomposed into representations of the Levi subalgebra:
\begin{equation}\label{GAlgDecLevi}
 \mathfrak{g}  = \mathcal{L} \oplus ... =   \bigoplus_{p} R_p^{\mathcal{L}}  \:,
\end{equation}
where the irreps $R_p^{\mathcal{L}}$ include the terms in the decomposition \eqref{PhiInL}. This will turn out useful in the following.

\subsubsection*{$A_3$ example}

\begin{figure}[H]
  \begin{center}
     \includegraphics[scale=0.30]{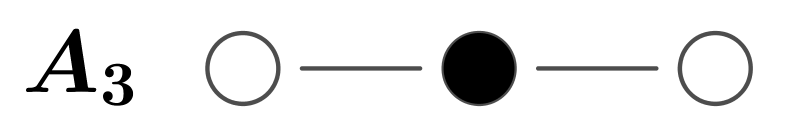}
  \end{center}
  \caption{Partial resolution of $A_3$}
  \label{figure 1}
  \end{figure}

Let us see how the choice of $\Phi$ works in a simple example: a family of deformed $A_3$ singularity where only the central simple root $\alpha_2$ is simultaneously resolved at the origin (see Figure \ref{figure 1}). 
We then have $\mathcal{H}=\langle\alpha_2^*\rangle$ and $\mathcal{W}'=\mathbb{Z}_2\times\mathbb{Z}_2$. The commutant of $\mathcal{H}$ is 
\begin{equation}
\mathcal{L} = A_1^{(1)} \oplus A_1^{(3)} \oplus \langle \alpha_2^* \rangle
\end{equation}
where $A_1^{(i)}$ ($i=1,3$) is generated by the triplet $e_{\pm\alpha_i},h_i$ normalized s.t.\ the commutation relations are $h_i=[e_{\alpha_i},e_{-\alpha_i}]$ and $[h_i,e_{\pm \alpha_i}]=\pm 2 e_{\pm \alpha_i}$. 
Following the prescription \eqref{recHiggsU} for each $A_1$ summand we have
\begin{equation}
\label{A1 Casimir}
\Phi|_{A_1^{(i)}} = \begin{pmatrix}
0 & 1 \\ \varrho_i & 0 
\end{pmatrix} = e_{\alpha_i}  + \varrho_i e_{-\alpha_i}  \qquad i=1,3
\end{equation}
where $\varrho_i$ ($i=1,3$) is the only Casimir of the $s\ell_2$ algebra $A_1^{(i)}$. 
$\Phi$ can also have a component along the Cartan $\alpha_2^*$, with generic coefficient $\varrho_2$.
$\varrho_1,\varrho_2,\varrho_3$ are the partial Casimirs relative to the chosen partial simultaneous resolution, i.e.\ they are invariant under the Weyl subgroup $\mathcal{W}'$. 
In matrix form, the chosen Higgs field vev is:
\begin{equation}\label{PhiA3Example}
\Phi = \begin{pmatrix}
\varrho_2 & 1 && \\
\varrho_1 & \varrho_2 && \\
&& -\varrho_2 & 1  \\
&& \varrho_3 & -\varrho_2  \\
\end{pmatrix}
\end{equation}
One can then take $\varrho_i=\varrho_i(w)$ in order to obtain a fibration over $\mathbb{C}_w$.

\subsection{The threefold equation from $\Phi$}\label{Sec:threefoldEq}

In order to obtain the explicit equation of the threefold we first derive the expression of the family of ADE surfaces fibered over $\mathfrak{t}/\mathcal{W}'$ with coordinates $\varrho_i$. We then let these coordinates depend on $w\in\mathbb{C}$. 

Katz and Morrison \cite{Katz:1992aa} developed a method for obtaining the hypersurface equation for an ADE family with a given simultaneous resolution, which we review in Appendix~\ref{App:Threefold}.
In the following, instead, we propose a novel method to read off the explicit equation of the simple flops in the ADE classification using the Higgs field $\Phi$.\footnote{Actually, our method works for any CY threefold derived from an ADE family.} Although the resulting expressions match those of \cite{Katz:1992aa}, this approach tackles the problem from a different perspective, unifying the physical description via the Higgs field $\Phi$ and the algebraic expressions of the deformed ADE singularities. 

We first consider the adjoint quotient map, that associates an element $\mathrm{g}$ of the ADE algebra $\mathfrak{g}$ of rank $r$ with $r$ independent polynomials $\chi_i$, whose value gives a point in $\mathfrak{t}/\mathcal{W}$:
\begin{equation}\label{adjoint quotient map}
\chi : \quad \mathfrak{g}\rightarrow \mathfrak{t}/\mathcal{W} : \mathrm{g}\mapsto (\chi_1(\mathrm{g}),\cdots,\chi_r(\mathrm{g})).
\end{equation}
The polynomials $\chi_i(\mathrm{g})$ are the Casimirs 
of $\mathfrak{g}$ and are defined in the following way.
By choosing a representation of the Lie algebra $\mathfrak{g}$, $\mathrm{g}$ can be put in a matrix form. Its Casimirs are then given by specific invariant polynomials of this matrix:
For the $A$ and $D$ series, the canonical Casimirs of $\mathrm{g}$ are (with $\mathrm{g}$ a matrix in the fundamental/vector representation):
\begin{equation}\label{AnDnCasimirs}
\begin{array}{c|l}
\boldsymbol{A_r} & \boldsymbol{D_r} \\
\hline
\chi^A_i(\mathrm{g})=\tfrac{1}{i+1}\text{Tr}(\mathrm{g}^{i+1}), \hspace{0.2cm} i=1,\ldots,r & \chi^D_i(\mathrm{g})=\text{Tr}(\mathrm{g}^{2i}),\hspace{0.2cm} i=1,\ldots,r-1 \\
 & \chi^D_r(\mathrm{g})=\text{Pfaff}(\mathrm{g}) \\
\end{array}
\end{equation}
For the exceptional algebras $E_r$, one takes $\mathrm{g}$ in the following representations:
$\boldsymbol{27}$ for $E_6$, $\boldsymbol{133}$ for $E_7$ and $ \boldsymbol{248} $ for $E_8$.
One then defines the Casimirs of $\mathrm{g}$ as\cite{Dixmier,Okubo}:
\begin{equation}\label{En casimirs}
\begin{array}{c|cl}
\boldsymbol{E_6} & \chi^{E_6}_i(\mathrm{g}) = \text{Tr}(\mathrm{g}^{k_i}) & \text{for }k_i=2,5,6,8,9,12  \\
\boldsymbol{E_7} & \chi^{E_7}_i(\mathrm{g}) = \text{Tr}(\mathrm{g}^{k_i}) & \text{for }k_i=2,6,8,10,12,14,18\\
\boldsymbol{E_8} & \chi^{E_8}_i(\mathrm{g}) = \text{Tr}(\mathrm{g}^{k_i}) & \text{for }k_i=2,8,12,14,18,20,24,30 \\
\end{array},
\end{equation}
and $i=1,...,r$.

We now  need to provide the definition of \textit{Slodowy slices}.
 Consider a nilpotent element~$\mathrm{x}\in\mathfrak{g}$  belonging to some nilpotent orbit $\mathcal{O}$: the Jacobson-Morozov theorem ensures that there exists a standard triple $\{\mathrm{x},\mathrm{y},\mathrm{h}\}$ of elements in $\mathfrak{g}$ satisfying the $\mathfrak{su}(2)$ algebra relations.\footnote{The triple related to $\mathrm{x}$ is unique up to conjugation.}
Now, we define the \emph{Slodowy slice} through the point $\mathrm{x}$ as those Lie algebra elements satisfying:
\begin{equation}\label{Slodowy definition}
\mathcal{S}_\mathrm{x}=\{\mathrm{z}\in \mathfrak{g}\hspace{0.2cm}|\hspace{0.2cm}[\mathrm{z}-\mathrm{x},\mathrm{y}]=0\}.
\end{equation}
For our purposes, we are  interested in the so-called \textit{subregular} nilpotent orbit $\mathcal{O}_{\text{subreg}}$, and in the Slodowy slice passing through it.\footnote{In general, the subregular nilpotent orbit of $\mathfrak{g}$ is defined as the only orbit of dimension $
    \text{dim}(\mathcal{O}_{\text{subreg}}) = \text{dim}(\mathfrak{g})-\text{rank}(\mathfrak{g})-2
$.}

 Let us immediately clarify the definitions given so far with a simple example.
Consider the $A_3$ Lie algebra, and a nilpotent element $\mathrm{x}$ lying in its subregular nilpotent orbit. The element $\mathrm{x}$ and its Slodowy slice are
 \begin{equation}\label{Slodowy A2}
     \mathrm{x} = \begin{pmatrix}
     0 & 1 & 0 & 0  \\
     0 & 0 & 1 & 0 \\
     0 & 0 & 0 & 0  \\     
     0 & 0 & 0 & 0  \\     
     \end{pmatrix} 
     \:,\qquad
\mathcal{S}_{\text{subreg}}=\left\{\left(\begin{array}{cccc}{a} & {1} & {0} & {0} \\ {b} & {a} & {1} & {0}  \\ {c} & {b} & {a} & {d} \\ {e} & {0} & {0} & {-3a} \end{array}\right) \hspace{0.2cm}\Bigg| \hspace{0.2cm} a, b, c, d, e \in \mathbb{C}\right\}.
\end{equation}

The crucial fact for us is the following:
Given an algebra $\mathfrak{g}$ in the ADE classification, and an element $\mathrm{x}\in\mathcal{O}_{\text{subreg}}$, the intersection of the Slodowy slice through $\mathrm{x}$ with the fiber of the adjoint quotient map \eqref{adjoint quotient map} is isomorphic to the versal deformation of the corresponding ADE singularity, namely:
\begin{equation}\label{ADE def isom}
\mathcal{S}_{\text{subreg}} \cap \chi^{-1}(\boldsymbol{u}) \cong \text{versal deformation of }\mathbb{C}^2/\Gamma_{\rm ADE},
\end{equation}
where $\boldsymbol{u}$ is a point in $\mathfrak{t}/\mathcal{W}\simeq \mathbb{C}^r$ with coordinates $u_i$.
In particular,  the intersection of $\mathcal{S}_{\text{subreg}}$ and the nilpotent cone\footnote{I.e. the set of all nilpotent elements.} $\mathcal{N}=\chi^{-1}(0)$ is isomorphic to the ADE singularity.

The isomorphism \eqref{ADE def isom} is telling us that the coefficients of the monomials in the versal deformation can be written in terms of the coordinates $u_i$ related to the polynomials $\chi_i$. 
In order to see this explicitly, let us come back to the $A_3$ example. In the $A$ cases, the Casimir invariants $\chi_i$ are given in \eqref{AnDnCasimirs}. Considering the elements in the Slodowy slice \eqref{Slodowy A2} through the subregular nilpotent orbit, we find:
\begin{eqnarray}\label{chi A3}
 \chi_1(\mathcal{S}_{\text{subreg}}) &=& 6a^2+2b,\nonumber\\
 \chi_2(\mathcal{S}_{\text{subreg}}) &=& -8a^3+4ab+c,\\
  \chi_3(\mathcal{S}_{\text{subreg}}) &=& 21a^4+6a^2b+3ac+2b^2+de\:.\nonumber
\end{eqnarray}
We now compute $ \mathcal{S}_{\text{subreg}} \cap \chi^{-1}(\boldsymbol{u})$:
\begin{equation}\label{A3eqslod}
\left\{ \begin{array}{l}
  6a^2+2b=u_1\\
-8a^3+4ab+c =u_2\\
 21a^4+6a^2b+3ac+2b^2+de=u_3\\
 \end{array}\right. \Rightarrow\,
 -cd=(3a)^4-u_1(3a)^2+u_2(3a)-u_3+\frac{u_1^2}{2}\:,
\end{equation}
that is the usual presentation of the deformed $A_3$ singularity: if we set $u=c,v=-d,z=3a$, we obtain
\begin{equation}\label{A3eqDef}
uv=z^4-u_1 z^2+u_2z-u_3+\frac{u_1^2}{2}
\end{equation}
that matches with \eqref{Anfamily} (up to an invertible redefinition of the coordinates in $\mathfrak{t}/\mathcal{W}$).\footnote{The match works for  $\sigma_2=u_1,\sigma_3=u_2,\sigma_4=u_3-u_1^2/2$.}

\

We now want to describe the ALE families of deformed ADE singularities over $\mathbb{C}_w$. We have constructed them by the choice of a Higgs field $\Phi(w)$, that through its Casimirs is telling us how the deformation is performed on top of each point of $\mathbb{C}_w$. In other words, take $w\in\mathbb{C}_w$; to see which is the deformed ALE surface over $w$, we pick $\Phi(w)$ and compute its Casimirs $\chi_i(\Phi)$. Their values select  a specific point $\boldsymbol{u}\in\mathfrak{t}/\mathcal{W}$ and then a specific deformed ALE surface (with precise volumes of the non-holomorphic spheres \eqref{volumes}). The equation of the threefold is obtained imposing the relations
\begin{equation}\label{3foldEqFromSlodowy}
    \chi_i(\mathcal{S}_{\text{subreg}})=\chi_i(\Phi(w)) \,,\qquad i=1,...,r\:,
\end{equation}
with $\chi_i(\mathrm{g})$ defined in \eqref{AnDnCasimirs} and \eqref{En casimirs}, and substituting the resulting $\chi_i(\mathcal{S}_{\text{subreg}})$ into the expressions defining $\mathcal{S}_{\text{subreg}}  \cap \chi^{-1}(\chi_i(\mathcal{S}_{\text{subreg}}))$.

In general, if we keep $\Phi(w)$ unspecified, \eqref{3foldEqFromSlodowy} gives us the versal deformations of the ADE singularity, with deformation parameters depending on the Casimirs of $\Phi$. We have worked out these relations for all the ADE algebras. For the $A_r$  and $D_r$ singularities, one obtains (up to coordinate redefinition) the compact and known form
\begin{equation}\label{An threefold}
A_r: \qquad x^2+y^2+\text{det}(z\mathbb{1}-\Phi)=0
\end{equation}
and
\begin{equation}\label{Dn threefold}
D_r: \qquad x^2+zy^2-\frac{\sqrt{\text{det}(z\mathbb{1}+\Phi^2)}-\text{Pfaff}^2(\Phi)}{z}+2y\hspace{0.1cm}\text{Pfaff}(\Phi)=0.
\end{equation}
For the $E_r$ singularities the expression of the deformation parameters $\epsilon_i,\tilde{\epsilon}_i,\hat{\epsilon}_i$ (appearing in \eqref{deformed ADE singularities}) in terms of the Casimirs \eqref{En casimirs} are given in Appendix~\ref{Appendix A}.

Hence, given a Higgs field $\Phi(w)$, one obtains the threefold equation by simply computing its Casimirs and then inserting them into the relations \eqref{An threefold} and \eqref{Dn threefold}  for the $A$ and $D$ cases, or using the formulae in Appendix~\ref{Appendix A} for the $E$ cases.

\subsubsection*{$A_3$ example}

In order to make the procedure explicit, let us apply it to the example of $A_3$.
Take the Higgs field \eqref{PhiA3Example} and compute its Casimirs:
\begin{eqnarray}
\chi_1(\Phi)&=&\varrho_1+\varrho_3+2\varrho_2^2 \nonumber \\
\chi_2(\Phi)&=&2\varrho_2(\varrho_1-\varrho_3)  \\
\chi_3(\Phi)&=&\frac12(\varrho_1+\varrho_3+2\varrho_2^2 )^2-(\varrho_2^2-\varrho_1)(\varrho_2^2-\varrho_3) 
 \nonumber 
\end{eqnarray}
We substitute $u_i$ with $\chi_i(\Phi)$ in \eqref{A3eqDef}, obtaining
\begin{equation}
uv=\left((z+\varrho_2)^2 -\varrho_1   \right)\left((z-\varrho_2)^2 -\varrho_3   \right)\:.
\end{equation}
One can check that the same equation can be obtained using directly \eqref{An threefold}.
Later we will be interested into the threefold given by $\varrho_1=-\varrho_3=w$ and $\varrho_2=0$, that leads to
$uv=z^4-w^2$.


\subsection{Gopakumar-Vafa invariants of $X$ from zero modes of $\Phi$}\label{Sec:ZeroModes}

In the non-compact Calabi-Yau threefolds studied in this paper, we have isolated singularities whose exceptional locus is given by a bunch of $\mathbb{P}^1$'s. These (genus-zero) holomorphic curves are rigid and the BPS M2-branes wrapped on them generate massless hypermultiplets in the 5d theory coming from reduction of M-theory in $X$. The genus-zero and degree $\mathbf{d}=(d_1,...,d_\ell)$ Gopakumar-Vafa invariants $n^{g=0}_{d_1,...,d_\ell}$ count such states. In particular, we say that
\begin{equation} \label{GVvsHypers}
n^{g=0}_{\mathbf{d}} = \# \mbox{ 5d hypers with charges $(d_1,...,d_\ell)$ under the  flavor group generated by }\mathcal{H} \:.
\end{equation}
These numbers are reproduced by counting the zero modes of the Higgs field $\Phi$, that are localized at $w=0$. 

Given a vev for $\Phi$, the zero modes are the deformation $\varphi \in \mathfrak{g}$ of the Higgs field up to the (linearized) gauge transformations
\begin{equation}\label{LinGaugeTr}
\delta_{g}\varphi = [\Phi,g] \qquad\mbox{with}\qquad g \in \mathfrak{g} \:.
\end{equation}





We take $\Phi=X_+ + w\,Y$ as explained at page \pageref{PhiXY}, where $X_+\in\mathcal{L}$  is in the principal nilpotent orbit of each $\mathcal{L}_h$.
We need to work out which components of the deformation $\varphi$ can be set to zero by a gauge transformation \eqref{LinGaugeTr}. One then tries to solve the equation
\begin{equation}\label{eqZeroModes}
  \varphi + \delta_{g} \varphi = 0 \,, \qquad\mbox{with}\qquad \delta_{g}\varphi = [X_++w\,Y,g] 
\end{equation}
with unknown $g \in \mathfrak{g}$. As we will see, at special points in $\mathbb{C}_w$, there can be components of $\varphi$ that cannot be gauge-fixed to zero: these directions in the Lie algebra $\mathfrak{g}$ support zero modes.

Since the irreducible representations $R^{\mathcal{L}}$ of $\mathcal{L}$ are invariant under the action of $\Phi$, we implement the decomposition \eqref{GAlgDecLevi} and we solve the equation \eqref{eqZeroModes}
in each representation $R^{\mathcal{L}}$ at a time, where now $g,\varphi\in R^{\mathcal{L}}\subset \mathfrak{g}$. We can write more explicitly the representation $R^{\mathcal{L}}$ of $\mathcal{L}=\mathcal{H}\oplus\mathcal{L}_1\oplus\mathcal{L}_2\oplus ...$ as
\begin{equation}
  R^{\mathcal{L}} = \left( \,\,\,R^{\mathcal{L}_1}\,\,,\,\, R^{\mathcal{L}_2}  \,\,,\,\, ...\,\,\, \right)_{q_1,...,q_\ell}
\end{equation}
where $R^{\mathcal{L}_h}$ is an irreducible representation of the simple summand $\mathcal{L}_h$ and $(q_1,...,q_\ell)$ are the charges under the $U(1)^\ell$ group generated by $\mathcal{H}$. 
If there are $n$ 5d modes in the representation $R^{\mathcal{L}}$, there will be other $n$ 5d modes in the conjugate representation $\bar{R}^{\mathcal{L}}$; together these generate $n$ massless hypermultiplets in the 5d theory localized at the singularity.
By using the correspondence \eqref{GVvsHypers}, we can then say that the GV invariant with degrees $(q_1,...,q_\ell)$ is
\begin{equation}
   n^{g=0}_{q_1,...,q_\ell} = \tfrac12 \cdot \# \,\mbox{localized zero-modes from }\, 
   R^{\mathcal{L}} \oplus \bar{R}^{\mathcal{L}} 
\end{equation}
with $R^{\mathcal{L}} = \left( R^{\mathcal{L}_1}, R^{\mathcal{L}_2}  , ... \right)_{q_1,...,q_\ell}$.

\

Let us describe our algorithm to compute the number and the charges of zero modes.
For each representation $R^{\mathcal{L}}$ of $\mathcal{L}$ with dimension $d_R$, 
we  choose a basis $e_1,...,e_{d_R}$ of $R^{\mathcal{L}}$. 
In this basis, the equation \eqref{eqZeroModes} becomes
\begin{equation}\label{eqZMAB}
 \left( A + w \, B\right) \bm \rho = -\bm \phi 
\end{equation}
where $\bm \rho$ and $\bm \phi$ are the $d_R$-column vectors of coefficients of $g$ and $\varphi$ in the given basis and $A,B$ are the constant $d_R\times d_R$ matrices representing the linear operators $X_+$ and $Y$ respectively.

If $A+w\,B$ is invertible, then there exists a vector $\bm \rho$ (i.e. a $g\in R^{\mathcal{L}}$) that completely gauge fixes $\varphi\in R^{\mathcal{L}}$  to zero at generic $w$. At the values of $w$ where the rank of $A+w\,B$ decreases, there will be vectors $\bm\phi$ that cannot be set to zero, leaving a zero mode localized at that points.

With the chosen $X_+$, we immediately see that such a special point is (by construction) the origin $w=0$. Here the matrix $A+w\,B$ reduces to the nilpotent matrix $A$, that has non-trivial kernel.\footnote{In particular, the kernel is spanned by the vectors $|j,j\rangle$, when writing $R^{\mathcal{L}}$ in terms of $sl_2$ representations, where $sl_2$ is generated by the Jacobson-Morozov standard triple associated with $X_+$.} 
In the following we only use the fact that $A$ has rank $r<d_R$; hence, our conclusions are valid also when $A$ is not necessarily nilpotent. What we are going to say of course applies also for a nilpotent $A$.

We choose the basis $e_1,...,e_{d_R}$ of $R^{\mathcal{L}}$ such that $A$ is in the Jordan form. If the rank of $A$ is $r$, we then have $d_R-r$ rows of zeros and $d_R-r$ columns of zeros. We can rearrange rows and columns such that $A$ takes the block diagonal form
\begin{equation}\label{AmatrixCanonicalForm}
A=\left(\begin{array}{l|l}
A_u & {\bm 0}_{r\times (d_R-r)}  \\ \hline {\bm 0}_{(d_R-r)\times r} & {\bm 0}_{(d_R-r)\times (d_R-r)} \\  
\end{array}\right) \:, \mbox{ with $A_u$ invertible.}
\end{equation}
Doing the same operations on $B$, we obtain
\begin{equation}
B=\left(\begin{array}{l|l}
B_u & B_r  \\ \hline B_l& B_d \\  
\end{array}\right) \:.
\end{equation}

The equations \eqref{eqZMAB} now read
\begin{equation}\label{eqZMBlockForm}
\left\{ \begin{array}{rcl}
(A_u  + w B_u ) \bm \rho_u + w B_r \bm \rho_d &=& - \bm\phi_u \\
  w B_l  \bm \rho_u + w B_d \bm \rho_d &=& - \bm\phi_d \\
\end{array}\right.
\end{equation}
Since $(A_u  + w B_u )$ is invertible (at least in the vicinity of $w=0$), from the first set of equations we see that we can always gauge fix the $\bm{\phi}_u$ components to zero,\footnote{These correspond to all states except $|j,-j\rangle$.} by setting 
\begin{equation}
\bm{\rho}_u = - (A_u + w\, B_u )^{-1} \left( \bm{\phi}_u+ w \, B_{r} \bm{\rho}_d \right)\:.
\end{equation}
Substituting in the second set of equations we obtain
\begin{equation}\label{eqZMBlockFormSecond}
\, w\, \left[   B_d - w \, B_l (A_u+ w\, B_u )^{-1} B_r  \right]  \bm{\rho}_d    =
 -  \bm{\phi}_d  + w\, B_l   (A_u+ w\, B_u )^{-1} \bm{\phi}_u \:.
\end{equation}
We see that  the components $\bm{\phi}_d$ cannot be fixed identically to zero: at $w=0$ there can be a remnant, i.e.\ a localized mode. Said differently, the best we can do is to cancel from $\bm{\phi}_d$ its dependence on $w$, leaving a constant entry (instead of a generic polynomial in $w$). This is possible for all components of $\bm\phi_d$ only when the matrix $B_d$ has maximal rank, i.e.\ rank equal to $d_R-r$. In this case, the number of zero modes is
$$
\#= d_R-r \:,
$$
because each component of $\bm\phi_d$ has now a constant entry, i.e. one degree of freedom.

If $B_d$ is not invertible, we can iterate what we have done so far, in the following way.
Let us define for simplicity $\bm\rho'\equiv \bm\rho_d$, $\bm \phi'_{\rm tot} \equiv \bm{\phi}_d  - w\, B_l   (A_u+ w\, B_u )^{-1} \bm{\phi}_u$, $A'\equiv B_d$ and $B'\equiv -B_l (A_u+ w\, B_u )^{-1} B_r$. We can decompose $\bm \phi'_{\rm tot} = \bm\phi'_0 + w \bm \phi'$, where $\bm\phi'_0$ is $\bm\phi'_{\rm tot}$ evaluated at $w=0$. We can then rewrite the equation \eqref{eqZMBlockFormSecond} as
\begin{equation}\label{eqZMABp}
 \left( A' + w \, B'\right) \bm \rho' = -\bm \phi'\:.
\end{equation}
This has the same form as \eqref{eqZMAB}, so we can again change the basis such that $A'\equiv B_d$ is in the Jordan form and write the equations in this basis. We will obtain a set of equations in the form \eqref{eqZMBlockForm} where we have to substitute $(A,B)_{u,l,r,d}\rightarrow (A',B')_{u,l,r,d}$ and $(\bm\rho,\bm\phi)\rightarrow (\bm\rho',\bm\phi')$. 

The matrix $A'$ will now have rank $r'<d_R-r$. There will then be $r'$ components of $\bm\phi'$ that can be gauge fixed to zero; we correspondingly have $r'$ zero modes along the corresponding components of $\bm\phi_d$. If the matrix $B'_d$ has maximal rank (i.e. $d_R-r-r'$), then the other $d_R-r-r'$ components of $\bm\phi_d$ will be of the form $a+bw$ and hence each hosts \emph{two} zero modes. In this case the number of zero modes is 
\begin{equation}\label{rank algorithm}
\#= r' + 2(d_R-r-r') \:.
\end{equation}

On the other hand, if $B'_d$ has rank $r''<d_R-r-r'$, then we have to iterate once more the algorithm above and, provided $B''_d$ has maximal rank (i.e. $d_R-r-r'-r''$) we obtain
$$
\#= r' + 2 r'' + 3 (d_R-r-r'-r'') \:.
$$
We now have the factor "$3$" because the $d_R-r-r'-r''$ directions of $\bm\phi_d$ are of the form $a+bw+cw^2$, i.e. they host \emph{three} zero modes each.

In conclusion, let us assume that the algorithm stops at the $N$-th step and let us call $r^{(k)}$ the rank of the matrix $A$ at the step $k$, then the number of zero modes is
\begin{equation}
\#_{\rm zero\,\ modes} =  \sum_{k=0}^N k\,r^{(k)} \qquad \mbox{with}\qquad \sum_{k=0}^Nr^{(k)}=d_R
\end{equation}
where $r^{(0)}=r$.

If there are other values of $w$, say $w=w_0$, where the rank of $A+wB$ is not maximal, one can shift $w\mapsto w+w_0$ ending out with the same situation as above, where the new $A$ is now $A+w_0B$. Applying the algorithm that we have just outlined, one computes the zero modes localized at $w=w_0$. In this case the matrix at $w=w_0$ is not necessarily nilpotent.

Notice that this algorithm could never end. This is the case for example when  the $A+wB$ matrix is identically zero at one step. The corresponding directions of $\varphi$ cannot be gauge fixed at any order in $w$, leaving a zero mode that lives in 7d.

\

In conclusion, in this section we have shown that one can  reduce the problem of finding the zero modes to a simple exercise in linear algebra. These computations are algorithmic and can be done by a calculator in a reasonable amount of time. In Appendix \ref{Appendix B} we describe the implementation of the algorithm in Mathematica, that we used for our computations. 

\section{Simple flops with $l=1,...,6$ and their GV invariants}
\label{Simple flop section}

One-parameter deformations of ADE surfaces admitting a small simultaneous resolution blowing up a single $\mathbb{P}^1$ are known as simple threefold flops. From a mathematical point of view, they can be classified according to a variety of invariants.

\indent A  rich classification of simple flops can be obtained employing the so-called \textit{length} invariant, first introduced in \cite{Kollar}, that we defined in Section~\ref{Sec:Intro}. 
It was proven (see \cite{Katz:1992aa}) that the length of a simple flop can only assume discrete values ranging from 1 to 6, and that examples of any length indeed exist. 

\begin{figure}[t]
  \begin{center}
     \includegraphics[scale=0.15]{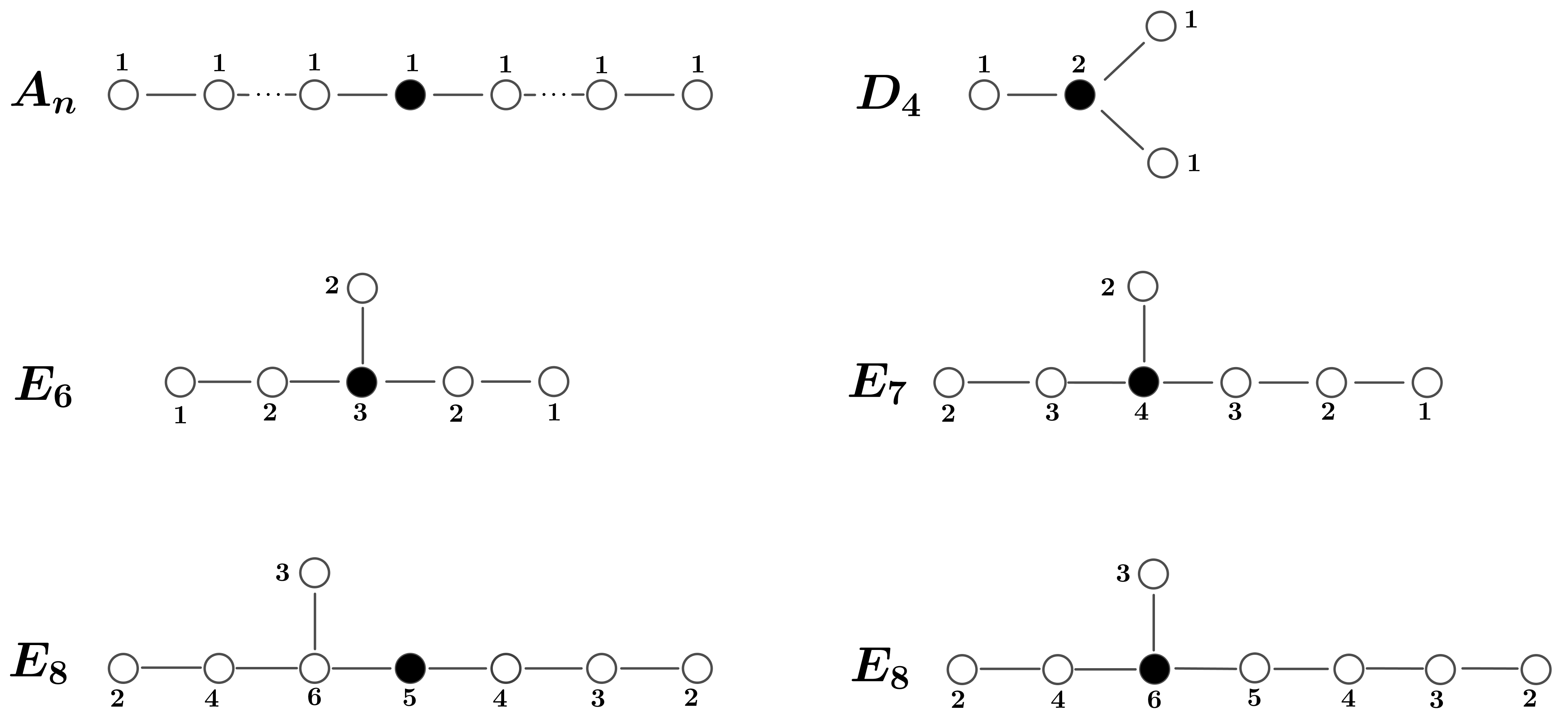}
  \end{center}
  \caption{ADE Dynkin diagrams and dual Coxeter labels of the nodes}\label{FigADESimpleFlops}
  \end{figure}

From a Lie-algebraic point of view, the length of a simple flop corresponds to the dual Coxeter label of the node of the Dynkin diagram that is being resolved by the small simultaneous resolution. Given an ADE algebra $\mathfrak{g}$ of rank $r$, a set of simple roots $\alpha_i$, with $i=1,\ldots r$, and the highest root $\theta$, the dual Coxeter label of a node is the multiplicity of the corresponding simple root in the decomposition of the highest root. In other words, given a node corresponding to a simple root $\alpha_{i_0}$ and the decomposition of the highest root
\begin{equation}
\theta = c_1\alpha_1 +\ldots+ c_{i_0}\alpha_{i_0}+\ldots c_r \alpha_r
\end{equation}
then $c_{i_0}$ is the dual Coxeter label of the node. As we will use this fact extensively in the explicit constructions of simple threefold flops of length up to 6, it is useful to report the Dynkin diagrams of all the ADE cases, along with the dual Coxeter labels of their nodes in Figure~\ref{FigADESimpleFlops}, 
where we have highlighted in black the nodes that are being resolved in the simple threefold flops that we will analyze in the following sections.\\
\indent The classification of simple threefold flops based on the length can be further refined introducing the Gopakumar-Vafa (GV) invariants \cite{Gopakumar:1998ii,Gopakumar:1998ki,Gopakumar:1998jq}. 
These invariants can be used to distinguish between simple flops of the same length.\footnote{Even though we will not use it in this paper, it is worth mentioning an even subtler invariant that can be associated to a simple flop, namely its contraction algebra. It has been proven \cite{BrownWemyss} that there exist simple flops with the same normal bundle, same length, same Gopakumar-Vafa invariants and \textit{different} contraction algebra. Physically, the contraction algebra can be understood, for example, as describing the quiver relations of the theory on a D3 brane in type IIB probing the singularity, and explicit constructions of contraction algebras at all lengths can be found in \cite{Karmazyn:2017aa}.}

In the following, we apply the method discussed in Section~\ref{Sec:3folds} to construct threefolds with a simple flop. The threefold will be obtained from a family of deformed ADE singularities in which only the black node in Figure~\ref{FigADESimpleFlops} is simultaneously resolved.
Let us call it $\alpha_c$. The subalgebra $\mathcal{H}$ is then generated by $\alpha_c^*$, i.e.
\begin{equation}
\mathcal{H} =\langle\alpha_c^*\rangle   \:,
\end{equation} 
and the Higgs field will correspondingly be chosen in the commutant $\mathcal{L}$ of $\mathcal{H}$, i.e.\ the Levi subalgebra corresponding to the chosen partial simultaneous resolution. From Figure~\ref{FigADESimpleFlops}, we see that the simple summands $\mathcal{L}_h$ of $\mathcal{L}$ are of $A$-type. The Higgs $\Phi$ restricted on these spaces is then of the form \eqref{recHiggsU}, and collecting the $\hat{\sigma}$'s from each summand $\mathcal{L}_h$ gives the partial Casimirs $\varrho_i$'s that parametrize the base $\mathfrak{t}/\mathcal{W}'$. The threefold is obtained by setting $\varrho_i=\varrho_i(w)$.

We will construct threefold with different values of \emph{length} from $1$ to $6$.
For each manifold we give the Higgs field $\Phi$ that produces the desired simple flop threefold $X$. This allows us to build the 5d theory realized from reducing M-theory on $X$. In particular  the flavor group will always be the $U(1)$ group generated by $\alpha_c^*$. The number of hypermultiplets and their charges under the $U(1)$ flavor group, namely the GV invariants of $X$ and their degrees, will be derived by counting the zero modes of $\Phi$.


\subsection{Simple flop with length $1$}

\subsubsection*{The simplest Example: the Conifold}

The Conifold threefold is given by
\begin{equation}\label{ConifoldEq}
u\,v= z^2-w^2\:.
\end{equation}
This is actually a family of deformed $A_1$ surfaces over $\mathbb{C}_w$, with the simultaneous resolution of the exceptional $\mathbb{P}^1$ at $w=0$. 

It can be constructed following the previous sections in the following way: $A_1$ has only one simple root $\alpha$. We require it to be blown up by the simultaneous resolution (the only other choice is to blow up no sphere, that would produce a non-singular threefold). The Levi subalgebra is now simply
\begin{equation}
 \mathcal{L}=\langle \alpha^*\rangle =\mathcal{H}.
\end{equation}
The partial Casimir is the coefficient $\varrho$ along the Cartan $\alpha^*$. Choosing $\varrho=w$, 
the Higgs field is
\begin{equation}
\Phi=\begin{pmatrix}
w&0\\0&-w\\
\end{pmatrix}
\end{equation}
and the threefold equation is easily checked to be \eqref{ConifoldEq}.

This is the simplest example of simple flop, where the flavor group (i.e.\ the preserved 7d gauge group) is $U(1)$.

\

{\bf Zero modes.} Notoriously, M-theory on the conifold gives a free 5d hypermultiplet (localized at $w=0$). This can be checked by computing the zero modes of $\Phi$. This computation has already been shown in \cite{Collinucci:2021ofd}, by explicitly using the linearized equations of motion in holomorphic gauge,\footnote{For the conifold, i.e.\ a family of deformed $A_1$, that computation was enough. However for more complicated algebras our method simplifies the calculations and make them more systematic.} as explained in \cite{Cecotti:2010bp}.
In order to illustrate the method outlined in Section~\ref{Sec:ZeroModes}, we apply it to the conifold case to reproduce the result of \cite{Collinucci:2021ofd}. 

The decomposition \eqref{GAlgDecLevi} of the $A_1$ algebra in representations of the Levi subalgebra $\mathcal{L}=\langle\alpha^*\rangle$ is
\begin{equation}
A_1={\bf 1}_0+{\bf 1}_++{\bf 1}_-.
\end{equation}
Let us consider each representation individually. Remember that the matrices $A$ and $B$ in Section~\ref{Sec:ZeroModes} are the restriction of $X_+$ and $Y$ on the considered representation, where $\Phi=X_++w\,Y$.
\begin{description}
\item ${\bf 1}_0$: $\Phi$ restricted to this representation is zero. Hence, the two `matrices' $A$ and $B$ vanish, nothing is gauge fixed and then there is one 7d mode.
\item ${\bf 1}_+$: $\Phi|_{ {\bf 1}_+}=2w$, so $A=0$ and it has rank zero, but $B=2$ has rank one; then $d_R-r=1-0=1$ and this gives one  localized mode at $w=0$.
\item ${\bf 1}_-$: $\Phi|_{ {\bf 1}_-}=-2w$, so $A=0$ and $B=-2$ that again gives a localized mode at $w=0$.
\end{description}
The two localized zero modes made up one hypermultiplet, as expected. Its charge under the flavor $U(1)$ can be easily read from the representation where the modes sit.
The zero mode analysis correctly reproduces the GV invariant of the conifold, that is $n_1^{g=0}=1$.

\subsubsection*{Threefolds with a simple flop of length $1$: generic case}

We now generalize the conifold case, by starting from the Lie algebra $A_{k-1}$. The simple roots are now $\alpha_1,...,\alpha_{k-1}$. We require that the only root that is simultaneously resolved in the threefold is $\alpha_p$ for a given choice of $p\in\{1,...,k-1\}$ (without loss of generality, we can take $p\geq \frac{k}{2}$). Consequently, we have $\mathcal{H}=\langle \alpha_p^* \rangle$. Its commutant is 
\begin{equation}
\mathcal{L} = A_{p-1}
\oplus A_{k-p-1}
\oplus \langle \alpha_p^* \rangle\:.
\end{equation} 
The Higgs field at $w=0$ is (in the principal nilpotent orbit when restricted on the simple summands of $\mathcal{L}$)
\begin{equation}
 X_+ =e_{\alpha_1}+ \dots + e_{\alpha_{p-1}}  + e_{\alpha_{p+1}} + \dots + e_{\alpha_{k-1}}\:.
\end{equation}
We choose the $w$-dependence of the partial Casimirs such that the Higgs restricted on each block is\footnote{A different choice would only complicate the equation of the three-fold, without changing its salient features.}
\be
\Phi|_{A_{p-1}}
=\left(\begin{array}{ccccc}
0 & 1 & 0 & \cdots & 0 \\
0 & 0 & 1 & 0 & 0\\
\vdots & 0 & \ddots & \ddots & 0  \\
0 & 0 & 0 & 0 & 1 \\
w & 0 & \cdots & 0& 0\\
\end{array} \right)
\qquad\mbox{and}\qquad
\Phi|_{A_{k-p-1}}
=\left(\begin{array}{ccccc}
0 & 1 & 0 & \cdots & 0 \\
0 & 0 & 1 & 0 & 0\\
\vdots & 0 & \ddots & \ddots & 0  \\
0 & 0 & 0 & 0 & 1 \\
-w & 0 & \cdots & 0& 0\\
\end{array} \right)\:.
\ee
This means that 
\begin{equation}
Y= e_{-\alpha_1-\alpha_2-... -\alpha_{p-1}} - e_{-\alpha_{p+1}-\alpha_{p+2}-... -\alpha_{2k-1}}\:.
\end{equation}
The equation of the threefolds is read form \eqref{An threefold}, by using the chosen $\Phi=X_++w\,Y$:
\begin{equation}
 u\,v=(z^{p}-w)(z^{k-p}+w) \:.
\end{equation}
When $k=2n$ is even and $p=n$, we have the \emph{Reid Pagoda} of degree $n$ (whose Dynkin diagram for the simultaneous resolution is depicted in Figure~\ref{FigADESimpleFlops}).

\

{\bf Zero modes.} Let us perform the zero mode computation in the case $p=n=2$, i.e. for the Reid Pagoda with degree $2$. The Higgs field is actually given by the $A_3$ example studied before (see \eqref{PhiA3Example}), where one chooses the following dependence of $\varrho_i$ on $w$:
$$
\varrho_1= w,\qquad \varrho_2=0,\qquad \varrho_3=-w\:.
$$
The $A_3$ algebra decomposes in the following way in representations of the Levi subalgebra $\mathcal{L}=s\ell_2^{(1)} \oplus s\ell_2^{(3)} \oplus \langle\alpha_2^*\rangle$:
\begin{equation}
   A_3 = ({\bf 3},{\bf 1})_0 \oplus  ({\bf 1},{\bf 3})_0 \oplus ({\bf 1},{\bf 1})_0 \oplus ({\bf 2},{\bf 2})_{+} \oplus ({\bf 2},{\bf 2})_{-}\:.
\end{equation}
Let us consider each Levi representation $R^{\mathcal{L}}$ individually.
\begin{description}
\item $({\bf 3},{\bf 1})_0$: the operator $X_+$ is represented in the basis $\{-e_{\alpha_1},\frac12 h_1,\frac12 e_{-\alpha_1}\}$ by the matrix 
\begin{equation}\label{pagoda irrep 31}
A_{({\bf 3},{\bf 1})_0} = \left( \begin{array}{c|cc}
0 & 1 & 0 \\  0 & 0 & 1 \\ \hline 0 & 0 & 0 \\
\end{array}\right) \:,
\end{equation}
that has rank $r=2$.
In the same basis $Y$ is represented by
\begin{equation}
B_{({\bf 3},{\bf 1})_0} = \left( \begin{array}{c|cc}
0 &  0 & 0 \\  2 & 0 & 0\\ \hline 0 & 2 & 0 \\
\end{array}\right)\:.
\end{equation}
We plug them into the expression \eqref{eqZMAB} and apply the algorithm: arranging the rows and columns to arrive to the expression \eqref{AmatrixCanonicalForm} is equivalent to taking $\bm \rho_u=(\rho_2,\rho_3)$, $\bm \rho_d=\rho_1$, $\bm \phi_u=(\phi_1,\phi_2)$ and  $\bm \phi_d=\phi_3$. We can then read
$$
B_u=\begin{pmatrix}
0&0\\0&0\\
\end{pmatrix},\qquad 
B_r=\begin{pmatrix}
0\\2\\
\end{pmatrix},\qquad 
B_l=\begin{pmatrix}
2&0\\
\end{pmatrix},\qquad 
B_d=0 \:.
$$
In particular $B_d-wB_l (A_u+ w\, B_u )^{-1} B_r$ vanishes identically. This means that at the second step $A'+wB'=0$ and the corresponding zero mode left by the rank 2 matrix A is not localized at any $w$. We have found a 7d zero mode.

\item $({\bf 1},{\bf 3})_0$: we obtain the same result as above, i.e.\ one 7d zero mode.

\item $({\bf 1},{\bf 1})_0$: $X_+$ and $Y$ vanish on this one-dimensional representation, leaving a 7d zero mode.

\item $({\bf 2},{\bf 2})_+$: the operator $X_+$ is represented in the basis $ \{e_{\alpha_1+\alpha_2+\alpha_3},e_{\alpha_1+\alpha_2}+e_{\alpha_2+\alpha_3},e_{\alpha_2},e_{\alpha_1+\alpha_2}-e_{\alpha_2+\alpha_3}\}$ by the matrix 
\begin{equation}
A_{({\bf 2},{\bf 2})_+} =  \left( \begin{array}{c|cc|c}
0 & 1 & 0 & 0 \\ 0 & 0 & 1 &  0 \\ \hline 0& 0 & 0 & 0 \\  0 & 0 & 0 & 0 \\
\end{array}\right)
\end{equation}
that has rank $r=2$.
In the same basis $Y$ is represented by
\begin{equation}
B_{({\bf 2},{\bf 2})_+} =  \left( \begin{array}{c|cc|c}
0 & 0 & 0 & 0 \\ 0 & 0 & 0 & 0 \\ \hline 0& 0 & 0 & 2 \\  1 & 0 & 0 & 0 \\
\end{array}\right)\:.
\end{equation}
We plug them into the expression \eqref{eqZMAB} and apply the algorithm: arranging the rows and columns to arrive to the expression \eqref{AmatrixCanonicalForm} is equivalent to taking $\bm \rho_u=(\rho_2,\rho_3)$, $\bm \rho_d=(\rho_1,\rho_4)$, $\bm \phi_u=(\phi_1,\phi_2)$ and  $\bm \phi_d=(\phi_3,\phi_4)$. We can then read
$$
B_u=\begin{pmatrix}
0&0\\0&0\\
\end{pmatrix},\qquad 
B_r=\begin{pmatrix}
0&0\\0&0\\
\end{pmatrix},\qquad 
B_l=\begin{pmatrix}
0&0\\0&0\\
\end{pmatrix},\qquad 
B_d=\begin{pmatrix}
0&2\\1&0\\
\end{pmatrix}\:.
$$
In particular $B_d$ has maximal rank, equal to $d_R-r=4-2=2$, where $r$ is the rank of $A$. This means that $\phi_d$ hosts two constant zero modes localized at $w=0$. These have charge~$+1$ with respect to the $U(1)$ generated by $\alpha_2^*$. Notice that det$(A+wB)=-2w^2$. Hence there are no other points in the base $\mathbb{C}_w$ that host localized zero modes.
\item $({\bf 2},{\bf 2})_-$: analogously to before, we have two zero modes localized at $w=0$ with charge~$-1$ under the preserved $U(1)$ group.
\end{description}

Hence, the number of localized modes at $w=0$ is $4$, that gives rise to two hypermultiplets with charge $1$ with respect to the $U(1)$ flavor group. This matches the results of \cite{Collinucci:2021wty,Collinucci:2021ofd}, that were obtained with a different procedure.

For generic $k$ and $p$, the zero mode counting proceeds analogously as for the Pagoda with $n=2$. The $A_{k-1}$ algebra decomposes in the following way in representations of the Levi subalgebra
\begin{equation}
   A_{k-1} = ({\boldsymbol{p^2-1}},{\bf 1})_0 \oplus  ({\bf 1},\boldsymbol{ (k-p)^2-1})_0 \oplus ({\bf 1},{\bf 1})_0 \oplus (\boldsymbol{ p},\boldsymbol{\overline{ k-p}})_{+} \oplus (\boldsymbol{ \bar{p}},\boldsymbol{ k-p})_{-}\:.
\end{equation}
The first three representations host 7d modes, but no localized one. 
Let us concentrate on the charged representation $(\boldsymbol{ p},\boldsymbol{\overline{ k-p}})_{+} $ of dimension $p(k-p)$. With the choice $p\geq\frac{k}{2}$, we have $p\geq k-p$. The matrix representing $X_+$ in this representation has kernel with dimension equal to $k-p$, then in our algorithm $r=(p-1)(k-p)$.
With a bit of work, one can check that $B_d$ has rank $k-p=d_R-r$, that gives then $k-p$ modes localized at $w=0$ with charge $+1$ with respect to the flavor $U(1)$. The other charge representation hosts again $k-p$ modes localized at $w=0$ and with charge $-1$. In total we then have $k-p$ charged hypermultiplets, i.e.\ the GV invariant is 
\begin{equation}
   n^{g=0}_1 = k-p\:.
\end{equation}

When $k=2n$ and $p=n$, we obtain $n$ hypers, that is the result we have found for the Reid's Pagodas in \cite{Collinucci:2021wty,Collinucci:2021ofd}, i.e. $n^{g=0}_1 = n$.

\subsection{Simple flop with length $2$}
\begin{figure}[H]
  \begin{center}
      \includegraphics[scale=0.35]{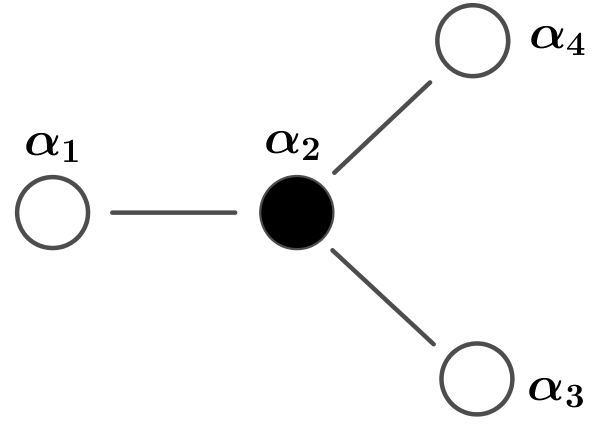}
  \end{center}
  \caption{$D_4$ Dynkin diagram}
   \label{D4dynkin}
  \end{figure}

In this section we consider a family of flops of length 2 arising from a $D_4$ singularity deformed over the $\mathbb{C}_w$ plane. The threefold is singular at the origin (where the fiber exhibits a $D_4$ singularity) and can only be partially resolved inflating a $\mathbb{P}^1$ corresponding to the central root of the $D_4$ Dynkin diagram. As we can see from Figure~\ref{FigADESimpleFlops} the central node has dual Coxeter label equal to 2, and thus its resolution yields a flop of length 2. In Figure~\ref{D4dynkin} we show our conventions for the labeling of the simple roots.

Since we wish to blow up only the central node, we have $\mathcal{H}=\langle \alpha_2^* \rangle$. The Levi subalgebra $\mathcal{L}$ commuting with $\mathcal{H}$ is:
\begin{equation}\label{D4 levi}
\mathcal{L} = A_1^{(1)}\oplus A_1^{(3)}\oplus A_1^{(4)}\oplus \langle \alpha_2^* \rangle,
\end{equation}
where the $A_1$ algebras correspond to the white ``tails'' in picture \ref{D4dynkin}, generated by the roots $\alpha_1$, $ \alpha_3$ and $ \alpha_4$ respectively.\\

Following the prescription \eqref{recHiggsU} for each $A_1$ summand we have
\begin{equation}\label{PhiA1i}
\Phi|_{A_1^{(i)}} = \begin{pmatrix}
0 & 1 \\ \varrho_i & 0 
\end{pmatrix} = e_{\alpha_i}  + \varrho_i e_{-\alpha_i}  \qquad i=1,3,4\:,
\end{equation}
where $\varrho_i$ ($i=1,3,4$) is the Casimir of the $s\ell_2$ algebra $A_1^{(i)}$. Moreover $\Phi$ can have a component along $\alpha_2^*$ with coefficient $\varrho_2$.
Although not necessary for the employment of our machinery, we report for the sake of visual clarity the explicit matrix form of the adjoint Higgs field corresponding to the choice \eqref{D4 levi}, employing the standard basis of \cite{Collingwood}:
\begin{equation}\label{Flop2Phi}
\Phi=\left(
\begin{array}{cccc|cccc}
\varrho_2  & 1 & 0 & 0 & 0 & 0 & 0 & 0 \\
 \varrho_1  & \varrho_2 & 0 & 0 & 0 & 0 & 0 & 0 \\
 0 & 0 & 0 & 1 & 0 & 0 & 0 & 1 \\
 0 & 0 & \varrho_3 & 0 & 0 & 0 & -1 & 0 \\
 \hline
 0 & 0 & 0 & 0  &  -\varrho_2 & -\varrho_1  & 0 & 0 \\
 0 & 0 & 0 & 0 & -1 & -\varrho_2 & 0 & 0 \\
 0 & 0 & 0 & -\varrho_4  & 0 & 0 & 0 & -\varrho_3  \\
 0 & 0 & \varrho_4 & 0 & 0 & 0 & -1 & 0 \\
\end{array}
\right).
\end{equation}

The threefold is found by imposing
\begin{equation}\label{Flop2Cutw}
\varrho_i(w)= w \,c_i(w) \qquad \mbox{for } i=1,2,3,4\:,
\end{equation}
where we take the $c_i(w)$'s such that $c_i(0)\neq 0$. Later we will simply choose  the $c_i(w)$'s to be constant in $w$.

The Higgs at the origin is then
\begin{equation}
X_+ = e_{\alpha_1}+e_{\alpha_3}+e_{\alpha_4},
\end{equation}
while $Y$ is
\begin{equation}
Y = c_1e_{-\alpha_1}+c_3e_{-\alpha_3}+c_4e_{-\alpha_4} + c_2 \langle\alpha_2^*\rangle\:.
\end{equation}

The threefold equation is simply obtained by taking the choice \eqref{Flop2Cutw} and the expression of $\Phi$ \eqref{Flop2Phi} and plugging them into the formula \eqref{Dn threefold}:\footnote{Notice that the threefold expression is not invariant under the exchange of $c_1, c_3$ and $c_4$, which are the Casimirs of the three $A_1$ tails: this can be overcome by a change of variables. In any case, the mode localization proceeds in a way that is invariant under the exchange of $c_1,c_3,c_4$.}
\begin{equation}
\begin{split}
& x^2+zy^2-z^3+w^2 z \left[c_1^2+c_3^2+c_4^2 +4 c_1 c_3+4c_1c_4-2c_3c_4 -2  c_2^2w (c_1-2 c_3-2c_4)+c_2^4 w^2\right]+\\
&-2 w^3 \left[c_1 \left( c_3^2+c_4^2 + c_1 c_3+c_1c_4-2c_3c_4\right)+c_2^2 w \left(c_3^2+c_4^2-2 c_1 c_3-2c_1c_4-2c_3c_4\right)+c_2^4 w^2 (c_3+c_4)\right]+\\
& -2 w z^2 \left(c_1+c_3+c_4+c_2^2 w\right)+ 2 w^2 y (c_3-c_4) \left(c_1-c_2^2 w\right)=0.
\end{split}\nonumber
\end{equation}

\vspace{-7mm}
Let's consider what happens when one of the $c_i$'s vanishes. If $c_2=0$, the preserved gauge group after Higgsing is $SU(2)$ instead of $U(1)$. This says that the ALE fiber has an $A_1$ singularity for all values of $w$, i.e.\ the threefold has a non-isolated singularity. 
If $c_i=0$ with $i=1,3,4$, then the preserved group is still $U(1)$. However, the threefold equation has an $A_1$ singularity for generic $w\in\mathbb{C}_w$: in fact, the threefold equation is the same one would obtain by taking $\Phi|_{A_1^{(i)}}$ identically zero (the equation is insensitive to the ``$1$'' in \eqref{PhiA1i}). Such a nilpotent vev for the Higgs field is called a T-brane \cite{{Cecotti:2010bp}}.

Since we want to consider isolated singularities (with a simple flop), avoiding T-brane configurations, we will take $c_i\neq 0$ $\forall i$.

\

{\bf Zero modes.} We now analyze the 5d zero modes arising from M-theory reduced on the flop of length 2 defined by \eqref{Flop2Cutw}. We keep the $c_i$'s as generic constants.

As in the case of the flops of length 1, the first step consists in determining the decomposition of the algebra $\mathfrak{g}=D_4$ into irreps of the Levi subalgebra \eqref{D4 levi}, obtaining:
\begin{equation}\label{D4 irreps}
D_4 = (\boldsymbol{3},\boldsymbol{1},\boldsymbol{1})_0\oplus (\boldsymbol{1},\boldsymbol{3},\boldsymbol{1})_0\oplus (\boldsymbol{1},\boldsymbol{1},\boldsymbol{3})_0 \oplus (\boldsymbol{2},\boldsymbol{2},\boldsymbol{2})_1 \oplus(\boldsymbol{2},\boldsymbol{2},\boldsymbol{2})_{-1} \oplus (\boldsymbol{1},\boldsymbol{1},\boldsymbol{1})_2\oplus (\boldsymbol{1},\boldsymbol{1},\boldsymbol{1})_{-2},
\end{equation}
where the numbers in parenthesis refer to representations of the three $A_1$ factors, and the subscript is the charge w.r.t.\ the Cartan $\langle \alpha_2^* \rangle$.
Let us examine the zero-mode content of the Levi representations in \eqref{D4 irreps} one by one:
\begin{description}
\item $(\boldsymbol{3},\boldsymbol{1},\boldsymbol{1})_0$: for this representation the story flows identically to the representation $(\boldsymbol{3},\boldsymbol{1})_0$ in the $A_3$ example, see \eqref{pagoda irrep 31}. The operator $X_+$ can be represented in the basis $\{-e_{\alpha_1},\frac{1}{2}h_1,\frac{1}{2}e_{-\alpha_1}\}$:
\begin{equation}
A_{(\boldsymbol{3},\boldsymbol{1},\boldsymbol{1})_0} = \left( \begin{array}{c|cc}
0 & 1 & 0 \\  0 & 0 & 1 \\ \hline 0 & 0 & 0 \\
\end{array}\right).
\end{equation}
Proceeding as in \eqref{pagoda irrep 31} it is easy to show that this representation does not host any localized 5d zero mode. The same holds for the representations $(\boldsymbol{1},\boldsymbol{3},\boldsymbol{1})_0$ and $(\boldsymbol{1},\boldsymbol{1},\boldsymbol{3})_0$.
\item $(\boldsymbol{1},\boldsymbol{1},\boldsymbol{1})_2$: $X_+$ is represented by a 1-dimensional matrix that, in the basis $e_{\alpha_1+2\alpha_2+\alpha_3+\alpha_4}$, reads
\begin{equation}
A_{(\boldsymbol{1},\boldsymbol{1},\boldsymbol{1})_2} = \left(0 \right).
\end{equation}
We also have:
\begin{equation}
B_{(\boldsymbol{1},\boldsymbol{1},\boldsymbol{1})_2} = 2 c_2.
\end{equation}
 As a result we find that $B$ has maximal rank, i.e.\ 1, and so we obtain \emph{one} localized 5d zero-mode with $U(1)$ charge 2.
Analogously, the representation $(\boldsymbol{1},\boldsymbol{1},\boldsymbol{1})_{-2}$ yields one 5d zero-mode of $U(1)$ charge $-2$.
\item $(\boldsymbol{2},\boldsymbol{2},\boldsymbol{2})_1$:  $X_+$, once put in Jordan form in an appropriate basis\footnote{The basis explicitly reads: $\{-e_{\alpha_{1}+\alpha_{2}}-e_{\alpha_{2}+\alpha_{3}},e_{\alpha_{1}+\alpha_{2}+\alpha_{4}}-e_{\alpha_{2}+\alpha_{3}+\alpha_{4}},\frac{2 e_{\alpha_{1}+\alpha_{2}}}{3}+\frac{e_{\alpha_{2}+\alpha_{3}}}{3}+\frac{e_{\alpha_{2}+\alpha_{4}}}{3},-\frac{1}{3} e_{\alpha_{1}+\alpha_{2}+\alpha_{3}}-\frac{1}{3} e_{\alpha_{1}+\alpha_{2}+\alpha_{4}}+\frac{2}{3} e_{\alpha_{2}+\alpha_{3}+\alpha_{4}},-6 e_{\alpha_{2}},-2 e_{\alpha_{1}+\alpha_{2}}+2 e_{\alpha_{2}+\alpha_{3}}+2 e_{\alpha_{2}+\alpha_{4}},e_{\alpha_{1}+\alpha_{2}+\alpha_{3}}+e_{\alpha_{1}+\alpha_{2}+\alpha_{4}}+e_{\alpha_{2}+\alpha_{3}+\alpha_{4}},e_{\alpha_{1}+\alpha_{2}+\alpha_{3}+\alpha_{4}} \}$.}, is represented as the 8-dimensional matrix
\begin{equation}\label{D4 A}
\scalemath{0.8}{
A_{(\boldsymbol{2},\boldsymbol{2},\boldsymbol{2})_1} = \left(
\begin{array}{cccccccc}
 0 & 1 & 0 & 0 & 0 & 0 & 0 & 0 \\
 0 & 0 & 0 & 0 & 0 & 0 & 0 & 0 \\
 0 & 0 & 0 & 1 & 0 & 0 & 0 & 0 \\
 0 & 0 & 0 & 0 & 0 & 0 & 0 & 0 \\
 0 & 0 & 0 & 0 & 0 & 1 & 0 & 0 \\
 0 & 0 & 0 & 0 & 0 & 0 & 1 & 0 \\
 0 & 0 & 0 & 0 & 0 & 0 & 0 & 1 \\
 0 & 0 & 0 & 0 & 0 & 0 & 0 & 0 \\
\end{array}
\right)},
\end{equation}
which has rank $r=5$. Using the same basis for $Y$ we get:
\begin{equation}
\scalemath{0.70}{
B_{(\boldsymbol{2},\boldsymbol{2},\boldsymbol{2})_1} =\left(
\begin{array}{cccccccc}
 c_2 & 0 & 0 & 0 & 6 c_4-6 c_3 & 0 & 0 & 0 \\
 c_1-c_3+c_4 & c_2 & \frac{2 (c_3-c_4)}{3} & 0 & 0 & 2 c_4-2 c_3 & 0 & 0 \\
 0 & 0 & c_2 & 0 & -6 (c_1+c_3-2 c_4) & 0 & 0 & 0 \\
 2 (c_1-c_3) & 0 & \frac{-c_1+5 c_3-c_4}{3}  & c_2 & 0 & -2 (c_1+c_3-2 c_4) & 0 & 0 \\
 0 & 0 & 0 & 0 & \rho & 0 & 0 & 0 \\
 0 & 0 & 0 & 0 & c_1+c_3+c_4 & c_2 & 0 & 0 \\
 \frac{c_3-c_1}{3} & 0 & \frac{2 c_1-c_3-c_4}{9} & 0 & 0 & \frac{4 (c_1+c_3+c_4)}{3} & c_2 & 0 \\
 0 & c_3-c_1 & 0 & \frac{2 c_1-c_3-c_4}{3}  & 0 & 0 & c_1+c_3+c_4 & c_2 \\
\end{array}
\right)\:.}
\end{equation}
Let us pause for a moment and use the results just found to prove that there are other isolated singularities in the threefold. In fact, these correspond to values of $w$ where 5d localized modes appear. This happens in the representation under study when the rank of $A_{(\boldsymbol{2},\boldsymbol{2},\boldsymbol{2})_1}+wB_{(\boldsymbol{2},\boldsymbol{2},\boldsymbol{2})_1}$  drops. Its determinant explicitly reads:
\begin{equation}\label{Det222}
\scalemath{0.9}{
\begin{split}
& \text{det}(A_{(\boldsymbol{2},\boldsymbol{2},\boldsymbol{2})_1}+wB_{(\boldsymbol{2},\boldsymbol{2},\boldsymbol{2})_1}) = w^4\left[(c_1^2+c_3^2+c_4^2-2 c_1 c_3-2 c_1 c_4-2 c_3 c_4)^2 +\right.\\
& \left.-4 c_2^2 w(c_1^3+c_3^3+c_4^3-c_1^2 c_3-c_1^2 c_4-c_3^2 c_1-c_3^2 c_4-c_4^2 c_1-c_4^2 c_3+10 c_1 c_3 c_4)+\right.\\
&\left. + 2 c_2^4 w^2(3 c_1^2+3 c_3^2+3 c_4^2+2 c_1 c_3+2 c_1 c_4+2 c_3 c_4)
-4 c_2^6 w^3 (c_1+c_3+c_4)+c_2^8 w^4\right].
\end{split}}
\end{equation}
It turns out that for generic $c_i$s the rank of $A_{(\boldsymbol{2},\boldsymbol{2},\boldsymbol{2})_1}+wB_{(\boldsymbol{2},\boldsymbol{2},\boldsymbol{2})_1}$ drops on top of $w=0$, as well as on further \textit{four} distinct points with non-zero $w$. It can be checked that these additional points correspond to conifold singularities far from the origin. In addition, if the condition
\begin{equation}\label{condition}
c_1^2+c_3^2+c_4^2 -2c_1c_3-2c_1c_4-2c_3c_4 = 0\:
\end{equation}
is satisfied, one of the additional singularities collides onto the origin: in this case, the rank of $A_{(\boldsymbol{2},\boldsymbol{2},\boldsymbol{2})_1}+wB_{(\boldsymbol{2},\boldsymbol{2},\boldsymbol{2})_1}$ drops on $w=0$ as well as on \textit{three} additional points outside the origin. This signals the appearance of further localized modes at $w=0$, coming from the conifold singularity that has collided onto the origin. We will explicitly check this claim momentarily, deriving again condition \eqref{condition}.

Rearranging rows and columns to get to the form \eqref{AmatrixCanonicalForm} we obtain:
\begin{equation}\label{D4 Bd}
\scalemath{0.8}{
\setlength{\jot}{16pt}
\begin{split}
& B_u = \left(
\begin{array}{ccccc}
 0 & 0 & 0 & 0 & 0 \\
 0 & 0 & 0 & 0 & 0 \\
 0 & 0 & 0 & 0 & 0 \\
 0 & 0 & c_2 & 0 & 0 \\
 0 & 0 & \frac{4(c_1+c_3+c_4) }{3} & c_2 & 0 \\
\end{array}
\right)\\
& B_r = \left(
\begin{array}{ccc}
 c_2 & 0 & 6 c_4-6 c_3 \\
 0 & c_2 & -6 (c_1+c_3-2 c_4) \\
 0 & 0 & c_2 \\
 0 & 0 & c_1+c_3+c_4 \\
 \frac{c_3-c_1}{3} & \frac{2 c_1-c_3-c_4}{9}  & 0 \\
\end{array}
\right)\\
& B_l=\left(
\begin{array}{ccccc}
 c_2 & 0 & 2 c_4 - 2 c_3 & 0 & 0 \\
 0 & c_2 & -2 (c_1+c_3-2c_4) & 0 & 0 \\
c_3-c_1 & \frac{2c_1-c_3-c_4}{3}  & 0 & c_1+c_3+c_4& c_2 \\
\end{array}
\right) \\
& B_d =\left(
\begin{array}{ccc}
 c_1-c_3+c_4 & \frac{2 (c_3-c_4)}{3} & 0 \\
 2 (c_1-c_3) & \frac{-c_1+5c_3-c_4}{3}  & 0 \\
 0 & 0 & 0 \\
\end{array}
\right) \\
\end{split}}
\end{equation}
Notice that the rank of $B_d$, which is surely non-maximal, depends on the precise choice of the partial Casimirs. It drops to one when its determinant is equal to zero. This happens when
\begin{equation}\label{rk1condci}
c_1^2+c_3^2+c_4^2 -2c_1c_3-2c_1c_4-2c_3c_4 = 0\:.
\end{equation}
Let us first examine the case in which the $c_i$'s are generic constants, i.e.\ $B_d$ has rank 2. Afterwards we see the case when $B_d$ has rank 1. Notice that $B_d$ cannot have rank zero, otherwise $c_1=c_3=c_4=0$, that we excluded.

\begin{itemize}
\item Let's take generic $c_i$'s such that $c_1^2+c_3^2+c_4^2 -2c_1c_3-2c_1c_4-2c_3c_4\neq 0$.
Renaming $A'\equiv B_d$ and $B'\equiv -B_l(A_u+wB_u)^{-1}B_r$ we can use equation \eqref{eqZMABp} to rerun the algorithm.  $A'$ is already in a form with a $2\times 2$ invertible block and all other elements equal to zero, i.e. $r'=2$. We can then immediately read $B_d'$ by computing the $(33)$ element of $B'$. It is
\begin{equation}
\scalemath{0.8}{
B_d' = 3 \left( c_1^2+c_3^2+c_4^2-2c_1c_3-2c_1c_4-2c_3c_4  \right)+\frac{10}{3} wc_2^2 (c_1+c_3+c_4)-c_2^4w^2},
\end{equation}
that has rank 1. As a result, according to \eqref{rank algorithm}, we find that the total number of zero modes is:
\begin{equation}\label{Bd zero modes}
\# = r'+2(d_R-r-r') = 2+2(8-5-2) = 4,
\end{equation}
where we recall that $d_R$ is the dimension of the representation, $r$ is the rank of \eqref{D4 A} and $r'$ is the rank of $A'$. The zero-modes have charge $+1$ with respect to the $U(1)$ generator.\\
Analogously, we find 4 localized zero-modes with charge $-1$ in the $(\boldsymbol{2},\boldsymbol{2},\boldsymbol{2})_{-1}$ representation.

\item When the $c_i$'s fulfill \eqref{rk1condci}, the rank of $B_d$ drops to 1. This produces a change in the zero-mode counting. We can parametrize a solution of  \eqref{rk1condci} in terms of two parameters $q_1,q_4$ as:
\begin{equation}
 c_1=q_1^2 ,\qquad c_3=(q_1+\varepsilon q_4)^2 ,\qquad c_4=q_4^2
\end{equation}
where $\varepsilon$ can take the values $\pm1$.
Now we have
\begin{equation}
A' = \left(
\begin{array}{ccc}
2q_1q_4 & \frac23q_1(q_1+2\varepsilon q_4) & 0 \\
 -2q_4(q_4+2\varepsilon q_1) & \frac23 ( 2q_1^2+5\varepsilon q_1q_4 +2q_4^2 ) & 0 \\
 0 & 0 & 0 \\
\end{array}
\right),
\end{equation}
When $q_1^2+\varepsilon q_1q_4+q_4^2\neq 0 $, the $2\times 2$ matrix is diagonalizable with the non-zero eigenvalue equal to $\frac{4}{3} \left(q_1^2+\varepsilon q_1q_4+q_4^2\right)$. The corresponding $B_d'$ is 
\begin{equation}
B_d' = 
\left(
\begin{array}{cc}
 c_2^2 & -  \frac{12 c_2 q_1 q_4 (q_1+\varepsilon  q_4)}{q_1^2+\varepsilon q_1 q_4+q_4^2} \\
 4 c_2 q_1q_4 (q_1+\varepsilon q_4)  & 0 \\
\end{array}
\right). 
\end{equation}
This matrix has rank less than two only when one of the $c_i$'s vanishes (and consequently the other two are equal to each other), that we excluded.

When $q_1^2+\varepsilon q_1q_4+q_4^2= 0 $ (i.e.\ all the eigenvalues vanish) the $2\times 2$ matrix has still rank 1 and the corresponding $B_d'$ is also forced to have rank 2 (for non-vanishing $c_i$'s).

We can finally count the localized zero-modes using formula \eqref{rank algorithm}, finding:
\begin{equation}
\# = r'+2(d_R-r-r') = 1+2(8-5-1) = 5.
\end{equation}
Notice that, with respect to the case \eqref{Bd zero modes} in which the Casimirs were totally generic, we have found an enhancement in the number of modes on a specific locus in the space of the partial Casimirs. This is the same locus where one conifold singularity that was at $w\neq 0$ collides onto the origin.

The representation $(\boldsymbol{2},\boldsymbol{2},\boldsymbol{2})_{-1}$ gives us further 5 zero-modes of charge $-1$.
\end{itemize}

\end{description}

Let us summarize our findings for the modes localized at $w=0$ for the simple flop of length 2 and partial Casimirs given by $\varrho_i(w)=w\,c_i$, with $c_i$ constants.
\begin{itemize}
\item For generic values of $c_i$'s, we get:
\begin{itemize}
\item[-] 8 modes with charge $\pm 1$,
\item[-] 2 modes with charge $\pm 2$.
\end{itemize} 
In terms of the GV invariants, this means 
\begin{equation}\label{GVFlop2-4}
n^{g=0}_1=4 \qquad\mbox{ and }\qquad n^{g=0}_2=1 \:.
\end{equation}
\item For $c_i$'s satisfying the constraint \eqref{rk1condci}, we get:
\begin{itemize}
\item[-] 10 modes with charge $\pm 1$,
\item[-] 2 modes with charge $\pm 2$.
\end{itemize} 
In terms of the GV invariants, this means 
\begin{equation}\label{GVFlop2-5}
n^{g=0}_1=5 \qquad\mbox{ and }\qquad n^{g=0}_2=1 \:.
\end{equation}

\end{itemize}

For the other (non-zero) values of $w$ where there are localized modes, we have conifold singularities and the flop is therefore not of length two: in fact, at these values of $w$ the $D_4$ is still deformed to a smaller singularity of $A$-type.\\

{\bf Non-constant $c_i$'s}. 
For simplicity, we have analyzed cases when the partial Casimirs $\varrho_i$ are just a constant $c_i$ multiplied by $w$. Of course, one can also let $c_i$ depend on $w$ and rerun the algorithm. 

One can in particular find the dependence of the $c_i(w)$'s such that the threefold $X$ has only one isolated singularity at the origin. An  easy solution is when
\begin{equation}
c_1=4a + b^2\,w \:,\qquad c_2= b \:,\qquad c_3=  c_4 = a  \:.
\end{equation}
One can check that for this choice the determinant \eqref{Det222} is equal to $-256a^3b^2w^5$, i.e. it vanishes only at $w=0$. The corresponding threefold has $n^{g=0}_1=5$ and $n^{g=0}_2=1$. For $a=-1/4$ and $b=1/2$ one actually recovers the Brown-Wemyss threefold \cite{BrownWemyss} in the form that appeared in \cite{Collinucci:2021ofd} (that has the expected GV invariants).


\subsection{Simple flop with length $3$}

\begin{figure}[H]
  \begin{center}
      \includegraphics[scale=0.28]{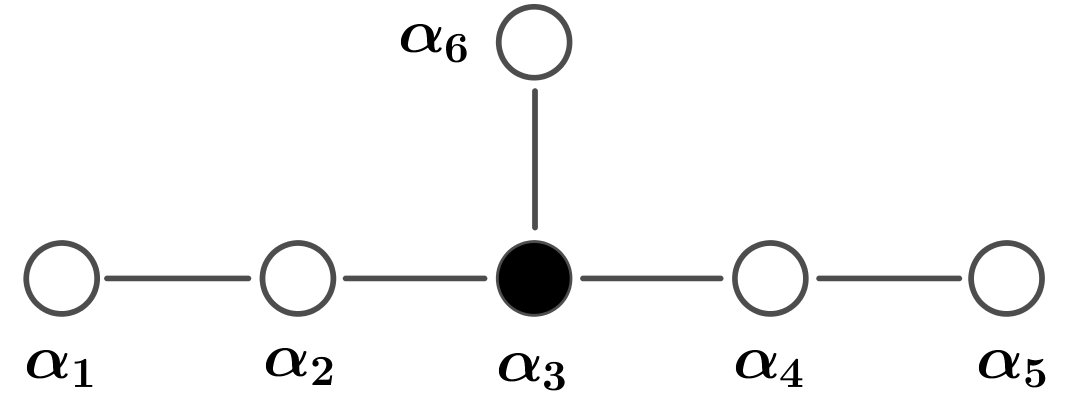}
  \end{center}
  \caption{$E_6$ Dynkin diagram, with the root blown up in the length three
    flop colored in black.}
   \label{E6dynkin}
  \end{figure}

In this section we engineer a threefold $X$ with a simple flop of length three. 
Analogously to the previous sections, we are going to define a suitable Higgs field, valued in the $E_6$ Lie
algebra, that generates  a family of deformed
$E_6$ surfaces with an $E_6$ singularity at $w =0$. The resolution of the isolated singularity in the threefold $X$ will blow-up only the trivalent node of the $E_6$ Dynkin
diagram (see Figure~\ref{FigADESimpleFlops}). To achieve this result, we pick the following Levi subalgebra 
\begin{equation}
  \label{levi length 3}
  \mathcal L = A^{(1,2)}_{2}\oplus A^{(4,5)}_{2} \oplus A_1^{(6)} \oplus \langle\alpha_3^*\rangle,
\end{equation}
where
 the factors
$A^{(i,j)}_{2}$ are associated, as subalgebras, to the roots $\alpha_{i}, \alpha_j$ of
the $E_6$ Dynkin diagram (we follow the labels in Figure~\ref{E6dynkin})  and $A_{1}^{(6)}$ is the algebra associated
to the root $\alpha_{6}$.

Again, we pick $X_{+}\equiv \Phi\rvert_{w =0}$ to be an element of the
principal nilpotent orbit of each simple factor of $\mathcal L$. The partial
Casimirs relative to $\mathcal L$ are the total Casimirs of each
simple factor of \eqref{levi length 3}, plus the coefficient along the Cartan element $\langle\alpha_3^*\rangle$. I.e. the generic $\Phi$ will be such that\footnote{To match the conventions of Section~\ref{Sec:ADEfamilies}, one takes $\{\varrho_i | \,\,i=1,...,6\}=\{\varrho_1^{(3)},\varrho_2^{(6)} \}\cup\{\varrho_3^{(i,j)},\varrho_2^{(i,j)}|\,\, (i,j)=(1,2),(4,5)  \}$. 
}
\begin{equation}
  \label{Higgs a2 tails e6}
  \Phi\rvert_{A_{2}^{(i,j)}}= 
    \left(
\begin{array}{ccc}
 0 & 1 & 0 \\
 0 & 0 & 1 \\
 \varrho^{(i,j)}_3 & \varrho^{(i,j)}_2 & 0 \\
\end{array}
\right)
=e_{\alpha_i}+e_{\alpha_j}+ \varrho^{(i,j)}_2e_{-\alpha_j} + \varrho^{(i,j)}_3[e_{-\alpha_j},e_{-\alpha_i}], 
\qquad i < j ,
\end{equation}
\begin{equation}
  \Phi\rvert_{A_{1}^{(6)}}= \left(
\begin{array}{ccc}
 0 & 1 \\
 \varrho^{(6)}_2 & 0 \\
\end{array}
\right)
=e_{\alpha_6} +\varrho^{(6)}_2e_{-\alpha_6} \qquad\mbox{and}\qquad
  \Phi\rvert_{\langle \alpha_3^* \rangle} = \varrho_1^{(3)} \langle \alpha_3^*\rangle	\:. \nonumber
\end{equation}
We now explicitly construct a threefold, by making the choice 
\begin{equation}
\begin{array}{lcl}
\varrho_1^{(3)} &=& w\, c_3\\
\varrho_{2}^{(6)} &=& w\, c_6\\
\varrho_{2}^{(1,2)} &=&0 \\
\varrho_{2}^{(4,5)} &=&0 \\
\varrho_{3}^{(1,2)} &=& w\, c_{12}\\
\varrho_{3}^{(4,5)} &=& w\, c_{45}\\
\end{array}
\end{equation}
with $c_3,c_6,c_{12},c_{45}$ constant numbers.

By plugging this choice into the Higgs field vev $\Phi$, and following the procedure described in Section~\ref{Sec:ConstructionMethod}, one obtains the threefold as an
hypersurface of $(x,y,z,w) \in \mathbb C^4$.\\
\indent As an example, if we pick $c_3=0,c_6=-3,c_{12}=1,c_{45}=-1$, one gets the following threefold, which is singular at the origin (as well as at other three points with non-zero $w$):
\begin{equation}
x^2+y^3+z^4+\frac{27 w^6}{32}+18 w^5+\left(12 w^3-\frac{27 w^4}{16}\right) y+2 \left(w^2-\frac{9 w^3}{8}\right) z^2+3 w y z^2=0\:.
\end{equation}
Via a change of coordinates, this exactly coincides with the length 3 threefold explicitly presented by \cite{Karmazyn:2017aa}.\\

{\bf Zero modes.} We now proceed (with the same procedure of the previous sections) to the mode counting. The branching of the adjoint
representation $\textbf{78}$ of $E_6$ w.r.t
$\mathcal L$ in \eqref{levi length 3} is given by\footnote{It can be better understood starting from the one of the maximal subalgebra
$A_2^{(1,2)}\oplus A_2^{(4,5)}\oplus A_{2}'$ (with $A_2'$ containing
$e_{\alpha_6}$):
$
  \textbf{78} = \textbf{(8,1,1)} \oplus \textbf{(1,8,1)} \oplus {\textbf{(1,1,8)}} \oplus {(\textbf{3},\overline{\textbf{3}},\textbf{3})} \oplus {(\overline{\textbf{3}},\textbf{3},\overline{\textbf{3}})}.
$
One then selects the subalgebra $A_1^{(6)}\subset A_{2}'$, and correspondingly
branches each term of the sum.}
\begin{eqnarray}
  \label{branching levi e6}
    \textbf{78}&=& \textbf{(8,1,1)}_0 \oplus \textbf{(1,8,1)}_0 \oplus \nonumber 
  {\textbf{(1,1,3)}_0 \oplus 
  (\textbf{1},\textbf{1},\textbf{2})_3 \oplus (\textbf{1},\textbf{1},\textbf{2})_{-3}
    \oplus
     (\textbf{1},\textbf{1},\textbf{1})_{0}} \oplus \nonumber \\
     &&   \oplus  {(\textbf{3},\overline{\textbf{3}},\textbf{2})_{1} \oplus (\textbf{3},\overline{\textbf{3}},\textbf{1})_{-2}}
        \oplus
        {
   (\overline{\textbf{3}},\textbf{3},\textbf{2})_{-1}\oplus
     (\overline{\textbf{3}},\textbf{3},\textbf{1})_{2}}\:, 
\end{eqnarray}
where the subscripts denote the charges under $\langle \alpha_3^* \rangle$.

For the $E$-cases the explicit computations done for length one and two become convoluted. We present here only the results. We
have worked out a Mathematica routine, presented in Appendix \ref{Appendix B}, that implements the algorithm described in Section~\ref{Sec:ZeroModes} and that can be used to check the results. Running this code for a generic choice of the parameters $c_6,c_3,c_{12},c_{45}$, we obtained, for each irreducible representation appearing in \eqref{branching levi e6}, the 5d modes shown in Table~\ref{modes e6}. In the table, we also write how many elements of the given representation support a mode localized in $\mathbb
C[w]/(w^k)$, for each $k$; we find that  $k \leq 2$.
\renewcommand{\arraystretch}{1.5}
\begin{table}[h!]
  \centering
  \begin{tabular}{ |p{2.cm}||p{2.cm}|p{2.cm}||p{2.cm}|}
    \hline
    $R^{\mathcal L}$ & $\mathbb C[w]/(w)$ & $\mathbb C[w]/(w^2)$& $\#_{\rm zero\, modes}$\\
    \hline
    $(\textbf{8},\textbf{1},\textbf{1})_0$ & 0     &  0           & 0   \\
    $(\textbf{1},\textbf{8},\textbf{1})_0$ & 0     &  0           & 0    \\
    $(\textbf{1},\textbf{1},\textbf{3})_{0}$ & 0     &  0   & 0 \\
    $(\textbf{1},\textbf{1},\textbf{1})_0$ & 0     &  0    & 0 \\
    \hline
    $(\textbf{3},\overline{\textbf{3}},\textbf{2})_{1}$ & 4     &  1    & 6   \\
    $(\overline{\textbf{3}},\textbf{3},\textbf{2})_{-1}$ & 4     &  1  & 6 \\
    $(\overline{\textbf{3}},\textbf{3},\textbf{1})_{2}$ & 3 &0  & 3 \\
    $(\textbf{3},\overline{\textbf{3}},\textbf{1})_{-2}$ & 3     &  0  & 3   \\
    $(\textbf{1},\textbf{1},\textbf{2})_{3}$ & 1     &  0           & 1    \\   
    $(\textbf{1},\textbf{1},\textbf{2})_{-3}$ & 1     &  0        & 1  \\
    \hline
  \end{tabular}
  \caption{5d modes for $E_6$ length three simple flop.}
  \label{modes e6}
\end{table}
We get a total of 20 5d modes: 
  \begin{itemize}
  \item one hyper with charge three, inside $(\textbf{1},\textbf{1},\textbf{2})_{3} \oplus
    (\textbf{1},\textbf{1},\textbf{2})_{-3}$;
  \item three hypers with charge two inside $(\overline{\textbf{3}},\textbf{3},\textbf{1})_{2} \oplus (\textbf{3},\overline{\textbf{3}},\textbf{1})_{-2}$;
  \item six hypers with charge one inside $(\textbf{3},\overline{\textbf{3}},\textbf{2})_{1}\oplus
    (\overline{\textbf{3}},\textbf{3}, \textbf{2})_{-1}$.
  \end{itemize}
In terms of the GV invariants, one then reads  
\begin{equation}\label{GVFlop3-genL3}
n^{g=0}_1=6\,, \qquad n^{g=0}_2=3 \qquad\mbox{ and }\qquad n^{g=0}_3=1 \:,
\end{equation}
which perfectly coincides with the results of \cite{Karmazyn:2017aa}.\\
\indent We can finally check whether there are special choices of the parameters
$c_{12}$, $c_{45}$, $c_{6},c_3$ for which the number of 5d modes localized at $w=0$ enhances.
A necessary condition for the enhancement of the number of modes is that
the rank of the matrix $B_d$  
drops for a special choice of the partial Casimirs. 
By explicit computation,
we find that the rank drops when $c_{12}=c_{45}$ or $c_6=0$. However,  these choices would create a non-isolated singularity.

\subsection{Simple flop with length $4$}

\begin{figure}[H]
  \begin{center}
      \includegraphics[scale=0.28]{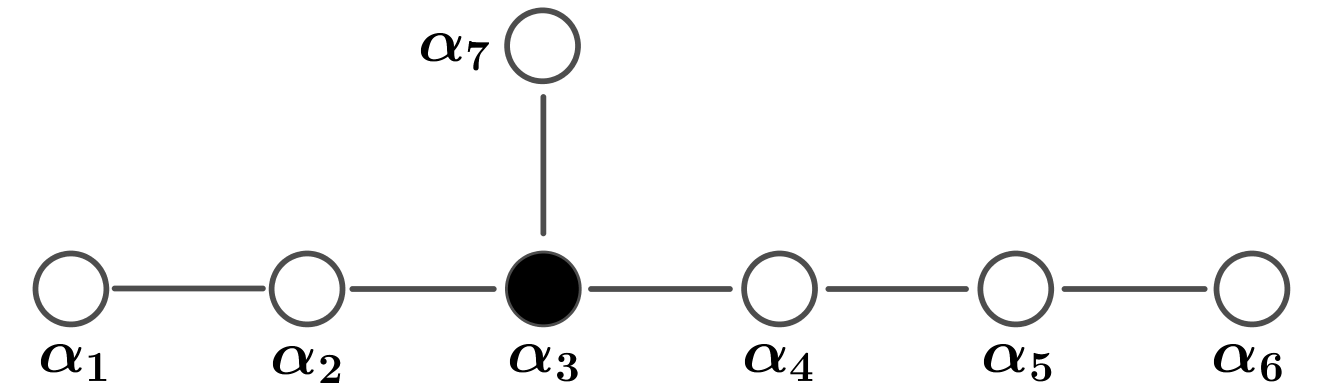}
  \end{center}
  \caption{$E_7$ Dynkin diagram, with the root blown up in the length four
    flop colored in black.}
   \label{E7dynkin}
  \end{figure}

In the following section we are going to engineer, by
means of a Higgs field $\Phi$ valued in the $E_{7}$ Lie algebra, a flop of length four. By
looking at the dual Coxeter labels of the $E_7$ Dynkin diagram in Figure~\ref{FigADESimpleFlops}, we see that the simultaneous resolution should involve the trivalent node. 
Analogously to the previous examples, this means that we have to pick the Higgs field in the
 Levi subalgebra
\begin{equation}
  \label{eq:73}
  \mathcal L \equiv A_{3}^{(4,5,6)} \oplus A_{2}^{(1,2)} \oplus A_{1}^{(7)}
  \oplus \langle \alpha_{3}^{*} \rangle,
\end{equation}
where the superscripts  refer to the roots of the $E_{7}$ Dynkin diagram
numbered as in the Figure~\ref{E7dynkin}, and $\alpha_{3}$ is the
trivalent root of $E_{7}$. 

Analogously to the $E_6$ case, we choose the Higgs field as follows:
\begin{equation}
     \Phi\rvert_{\langle\alpha_{3}^*\rangle}=c_3w \langle
\alpha_{3}^*\rangle
\end{equation}
and
\begin{eqnarray}
  \label{Higgs a2 tails e7}
      \Phi\rvert_{A_{1}^{(7)}}&=& 
    \left(
\begin{array}{cc}
 0 & 1  \\
 c_7w & 0  \\
\end{array}
\right)
=e_{\alpha_7}
  +c_7w\,e_{-\alpha_7}, \nonumber \\
    \Phi\rvert_{A_{2}^{(1,2)}}&=& 
    \left(
\begin{array}{ccc}
 0 & 1 & 0 \\
 0 & 0 & 1 \\
c_{12}w & 0 & 0 \\
\end{array}
\right)
=e_{\alpha_1}+e_{\alpha_2}+ c_{12}w\,[e_{-\alpha_1},e_{-\alpha_2}],
  \nonumber \\
  \Phi\rvert_{A_{3}^{(4,5,6)}}&=& 
    \left(
\begin{array}{cccc}
 0 & 1 & 0 & 0 \\
 0 & 0 & 1 & 0 \\
  0 & 0 & 0 & 1 \\
  c_{456}w  & 0 & 0 & 0 \\
\end{array}
\right)
=e_{\alpha_4}+e_{\alpha_5}+ e_{\alpha_6}+c_{456}w \Big[[e_{-\alpha_4},e_{-\alpha_5}],e_{-\alpha_{6}}\Big].
  \nonumber 
\end{eqnarray}
The corresponding threefold is a hypersurface in $\mathbb{C}^4$, that is a family of deformed $E_{7}$ singularities over $\mathbb{C}_w$.
To make the equation of the threefold more readable, we set the parameters to specific values, picking $c_3=0,c_7 =3,c_{12}=\frac{1}{2},c_{456}=-\frac{1}{2}$, obtaining
\begin{equation}
\begin{split}
& x^2-y^3+y z^3+3 w y^2 z+y^2 \frac{81w^2}{16}-y z\frac{w^2}{12}+z^2\frac{5 w^3}{8} -y\frac{w^3}{108}+z\frac{w^4}{3}+\frac{w^5}{144}=0.
\end{split}
\end{equation}
where we neglected terms of high degree, irrelevant for the singularity at $w=0$.\\

{\bf Zero modes.} We  now proceed with the modes
counting. We will again perform the gauge-fixing separately in each irreducible
representation of the branching of the adjoint representation
$\textbf{133}$ of $E_{7}$ under the subalgebra $\mathcal
L$:\footnote{The first entry of each summand is a representation of $A_{3}^{(4,5,6)}$, the second one is a representation of $A_{2}^{(1,2)}$, and the third on a representation of $A_{1}^{(7)}$. The subscript is the charge under $\langle \alpha_3^* \rangle$.}
\begin{eqnarray}
  \label{eq:176}
  \textbf{133} &=&
  (\textbf{15},\textbf{1},\textbf{1})_{0} \oplus
  (\textbf{1},\textbf{8},\textbf{1})_{0} \oplus
                   (\textbf{1},\textbf{1},\textbf{3})_{0} \oplus    (\textbf{1},\textbf{1},\textbf{1})_{0} \oplus \nonumber \\
               &&
                  (\overline{\textbf{4}},\textbf{3},\textbf{2})_{-1}\oplus
                  (\textbf{4},\overline{\textbf{3}},\textbf{2})_{1}
                  \oplus 
  (\textbf{6},\overline{\textbf{3}},\textbf{1})_{-2} \oplus
  (\textbf{6},\textbf{3},\textbf{1})_{2} \oplus 
  \nonumber \\
  &&
     (\textbf{4},\textbf{1},\textbf{2})_{-3} \oplus
     (\overline{\textbf{4}},\textbf{1},\textbf{2})_{3} \oplus   (\textbf{1},\textbf{3},\textbf{1})_{-4} \oplus
  (\textbf{1},\overline{\textbf{3}},\textbf{1})_{4}
.
\end{eqnarray}

Running the Mathematica routine described in Appendix \ref{Appendix B}, we find the results displayed in table \ref{modes e7}. As in the $E_{6}$ case,
there are no five-dimensional modes localized in $\mathbb C[w]/(w^k)$, with
$k > 2$.
\renewcommand{\arraystretch}{1.5}
\begin{table}[h!]
  \centering
  \begin{tabular}{ |p{2.cm}||p{2.cm}|p{2.cm}||p{2.cm}|}
    \hline
    $R^{\mathcal L}$ & $\mathbb C[w]/(w)$ & $\mathbb C[w]/(w^2)$ & $\#_{\rm zero\,modes}$ \\
    \hline
    $(\textbf{15},\textbf{1},\textbf{1})_0$ & 0     &  0          &0    \\
    $(\textbf{1},\textbf{8},\textbf{1})_0$ & 0     &  0         &0      \\
    $(\textbf{1},\textbf{1},\textbf{3})_{0}$ & 0     &  0       &0        \\
    $(\textbf{1},\textbf{1},\textbf{1})_{0}$ & 0     &  0        &0       \\
    \hline
    $(\overline{\textbf{4}},\textbf{3},\textbf{2})_{-1}$ & 6     &
                                                                           0   & 6 \\
    $(\textbf{4},\overline{\textbf{3}},\textbf{2})_{1}$ & 6     &  0   &6 \\
    $(\textbf{6},\overline{\textbf{3}},\textbf{1})_{-2}$ & 3 &1  &5 \\
    $(\textbf{6},\textbf{3},\textbf{1})_{2}$ & 3     &  1  &5    \\
    $(\textbf{4},\textbf{1},\textbf{2})_{-3}$ & 2     &  0 &2 \\
    $(\overline{\textbf{4}},\textbf{1},\textbf{2})_{3}$ & 2
                                          &  0   &2  \\
    $(\textbf{1},\textbf{3},\textbf{1})_{-4}$ & 1     &  0 &1 \\
    $(\textbf{1},\overline{\textbf{3}},\textbf{1})_{4}$ & 1     &  0   &1  \\
    \hline
  \end{tabular}
  \caption{five-dimensional modes for $E_7$ length four simple flop.}
  \label{modes e7}
\end{table}
In total, we find 28 modes localized at $w=0$:
  \begin{itemize}
  \item one hyper with charge four, inside $(\textbf{1},\textbf{3},\textbf{1})_{-4} \oplus
    (\textbf{1},\overline{\textbf{3}},\textbf{1})_{4}$;
  \item two hypers with charge three inside $(\textbf{4},\textbf{1},\textbf{2})_{-3} \oplus (\overline{\textbf{4}},\textbf{1},\textbf{2})_{3}$;
  \item five hypers with charge two inside $(\textbf{6},\overline{\textbf{3}},\textbf{1})_{-2}\oplus
    (\textbf{6},\textbf{3},\textbf{1})_{2}$.
  \item six hypers with charge one inside $(\overline{\textbf{4}},\textbf{3},\textbf{2})_{-1}\oplus
    (\textbf{4},\overline{\textbf{3}},\textbf{2})_{1}$.
  \end{itemize}
In terms of the GV invariants, one then reads  
\begin{equation}\label{GVFlop3-genL4}
n^{g=0}_1=6\,, \qquad n^{g=0}_2=5\,, \qquad n^{g=0}_3=2 \qquad\mbox{ and }\qquad n^{g=0}_4=1 \:.
\end{equation}

Finally, 
we find (as in the $E_6$ case) that no particular choice of the constants $c_i$  can  enhance the number of zero modes at $w=0$ (without generating non-isolated singularities).

\subsection{Simple flop with length $5$}
\label{Length five flop section}

\begin{figure}[H]
  \begin{center}
      \includegraphics[scale=0.28]{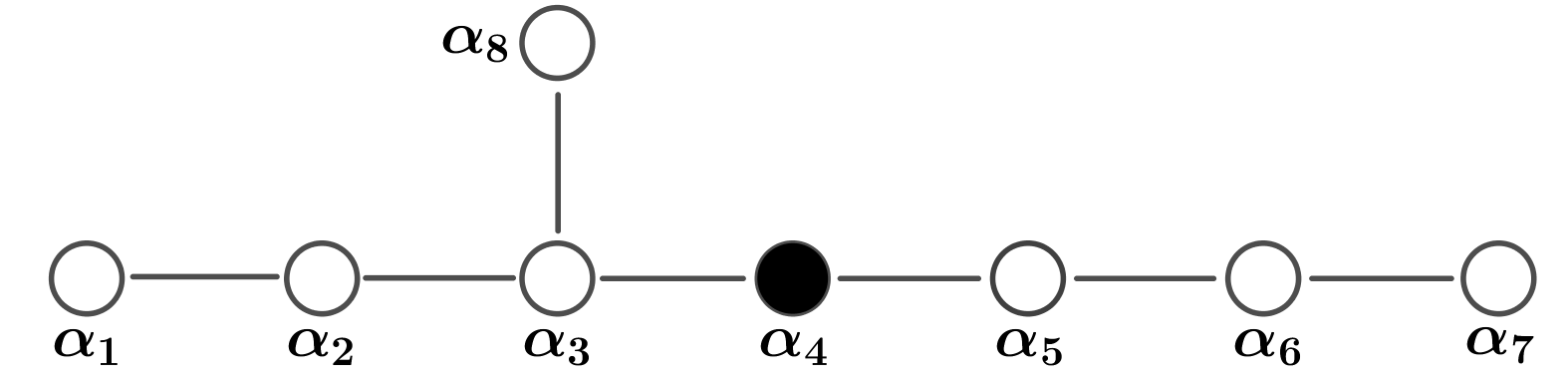}
  \end{center}
  \caption{$E_8$ Dynkin diagram, with the root blown up in the length five
    flop colored in black.}
   \label{E8dynkinlength5}
  \end{figure}

A flop with length 5 is obtained from an $E_8$ family over $\mathbb{C}_w$. The node that should be simultaneously resolved at $w=0$ is depicted in Figure~\ref{E8dynkinlength5}.

We then have $\mathcal{H}=\langle\alpha_4^* \rangle$ and 
\begin{equation}
  \label{eq:164}
  \mathcal L =  A_{3}^{(5,6,7)}\oplus A_{4}^{(1,2,3,8)} \oplus \langle \alpha_4^*\rangle.  
\end{equation}
We make the simple choice
\begin{eqnarray}\label{Higgs tails length five}
\Phi\rvert_{\langle\alpha_4^*\rangle} &=& c_4w\, \langle\alpha_4^*\rangle \nonumber\\
      \Phi\rvert_{A_{3}^{(5,6,7)}}&=&
    \left(
\begin{array}{cccc}
  0 & 1 & 0 & 0  \\
  0 & 0 & 1 & 0 \\
  0 & 0 & 0 & 1 \\
  c_{567}w & 0 & 0 & 0
\end{array}
\right)
=e_{\alpha_5} + e_{\alpha_6} + e_{\alpha_7} + 
   c_{567}w\left[[e_{-\alpha_5},e_{-\alpha_6}],e_{-\alpha_7}\right]\nonumber \\
    \Phi\rvert_{A_{4}^{(1,2,3,8)}}&=& 
\scalemath{0.8}{\left(  \begin{array}{ccccc}
 0 & 1 & 0 & 0 & 0 \\
 0 & 0 & 1 & 0 & 0 \\
 0 & 0 & 0 & 1 & 0 \\
 0 & 0 & 0 & 0 & 1 \\
 c_{1238}w& 0 & 0 & 0 & 0 \\
\end{array}
\right) =e_{\alpha_1}+e_{\alpha_2}+ e_{\alpha_3}+e_{\alpha_8}-   
  c_{1238}w\left[
  \Big[
  [e_{-\alpha_1},e_{-\alpha_2}],e_{-\alpha_3}
  \right],
  e_{-\alpha_8}
  \Big]}  \nonumber
\end{eqnarray}
with constant $c$'s.
We obtain our threefold as an
hypersurface in $\mathbb C^4$. To make the equation more readable, we pick explicit values for the parameters, setting $c_4=0,c_{567}=1,c_{1238}=-1$:
\begin{equation}
x^2+y^3+z^5+w^7+\frac{w^6}{864}-\frac{23 w^5 z}{36}-\frac{w^4 y}{48}-\frac{187 w^4 z^2}{36}-\frac{13}{3} w^3 y z-\frac{2 w^3 z^3}{27}-\frac{1}{3} w^2 y z^2=0 \:.
\end{equation}

{\bf Zero modes.} We can explicitly perform the branching of the adjoint representation
$\textbf{248}$ of $E_8$  under the
chosen $\mathcal L$:\footnote{The first number denotes the dimension of the
representation of $A_{3}^{(5,6,7)}$, the second under $A_{4}^{(1,2,3,8)}$ and the subscript is the charge under the Cartan $\alpha_4^*$.}
\begin{eqnarray}
  \label{branching length five}
  \textbf{248} &=& (\textbf{1},\textbf{24})_0 \oplus
  (\textbf{15},\textbf{1})_0 \oplus (\textbf{1},\textbf{1})_0 \oplus
                   \nonumber \\
  && (\textbf{4},\overline{\textbf{10}})_{1}\oplus
     (\overline{\textbf{4}},\textbf{10})_{-1} \oplus
     (\textbf{6},\textbf{5})_2 \oplus
     (\textbf{6},\overline{\textbf{5}})_{-2} \oplus  \nonumber \\
  && (\overline{\textbf{4}},\overline{\textbf{5}})_{3} \oplus
     (\textbf{4},\textbf{5})_{-3} \oplus
     (\textbf{1},\textbf{10})_{4} \oplus
     (\textbf{1},\overline{\textbf{10}})_{-4} \oplus \nonumber \\
  && (\textbf{4},\textbf{1})_{5}\oplus (\overline{\textbf{4}},\textbf{1})_{-5}.
\end{eqnarray}
The result of the zero mode counting is displayed in Table~\ref{modes e8 length five}.
There are no modes localized in
$\mathbb C[w]/(w^k),$ with $k > 2$. We find $48$ modes localized at $w=0$: 
\renewcommand{\arraystretch}{1.5}
\begin{table}[t]
  \centering
  \begin{tabular}{ |p{2.cm}||p{2.cm}|p{2.cm}||p{2.cm}|}
    \hline
    $R^{\mathcal L}$ & $\mathbb C[w]/(w)$ & $\mathbb C[w]/(w^2)$ &$\#_{\rm zero \, modes}$\\
    \hline
    $(\textbf{1},\textbf{24})_0$ & 0     &  0  &  0              \\
    $(\textbf{15},\textbf{1})_0$ & 0     &  0      &  0         \\
    $(\textbf{1},\textbf{1})_0$ & 0     &  0        &  0       \\
    \hline
    $(\textbf{4},\overline{\textbf{10}})_{1}$ & 6     &  1        &  8       \\
    $(\overline{\textbf{4}},\textbf{10})_{-1}$ & 6     & 1  &  8 \\
    \hline
    $(\textbf{6},\textbf{5})_2$ & 6     &
                                                                           0   &  6 \\
    $(\textbf{6},\overline{\textbf{5}})_{-2}$ & 6     &      0  &  6  \\
    \hline
    $(\overline{\textbf{4}},\overline{\textbf{5}})_{3}$ & 4 &0  &  4 \\
    $(\textbf{4},\textbf{5})_{-3}$ & 4    &  0   &  4  \\
    \hline                                                             
    $(\textbf{1},\textbf{10})_{4}$ & 2     &  0  &  2 \\
    $(\textbf{1},\overline{\textbf{10}})_{-4}$ & 2 &  0  &  2   \\
    \hline
    $(\textbf{4},\textbf{1})_{5}$ & 1     &  0  &  1\\
    $(\overline{\textbf{4}},\textbf{1})_{-5}$ & 1     &  0  &  1     \\
    \hline
  \end{tabular}
  \caption{five-dimensional modes for $E_8$ length five simple flop.}
  \label{modes e8 length five}
\end{table}
  \begin{itemize}
  \item one hyper with charge five, inside $(\textbf{4},\textbf{1})_{5} \oplus
  (\overline{\textbf{4}},\textbf{1})_{-5}$;
  \item two hyper with charge four, inside $(\textbf{1},\textbf{10})_{4}\oplus
    (\textbf{1},\overline{\textbf{10}})_{-4}$;
  \item four hypers with charge three inside $(\overline{\textbf{4}},\overline{\textbf{5}})_{3} \oplus (\textbf{4},\textbf{5})_{-3}$;
  \item six hypers with charge two inside $(\textbf{6},\textbf{5})_2\oplus
    (\textbf{6},\overline{\textbf{5}})_{-2}$;
  \item eight hypers with charge one inside $(\textbf{4},\overline{\textbf{10}})_{1}\oplus
    (\overline{\textbf{4}},\textbf{10})_{-1}$.
  \end{itemize}
In terms of the GV invariants, one then reads  
\begin{equation}\label{GVFlop3-genL5}
n^{g=0}_1=8\,, \qquad n^{g=0}_2=6\,, \qquad n^{g=0}_3=4\,, \qquad n^{g=0}_4=2 \qquad\mbox{ and }\qquad n^{g=0}_5=1 \:.
\end{equation}

Again, we notice that we can not enhance the
number of zero-modes at $w=0$ without generating a non-isolated singularity.

\subsection{Simple flop with length $6$}

\begin{figure}[H]
  \begin{center}
      \includegraphics[scale=0.28]{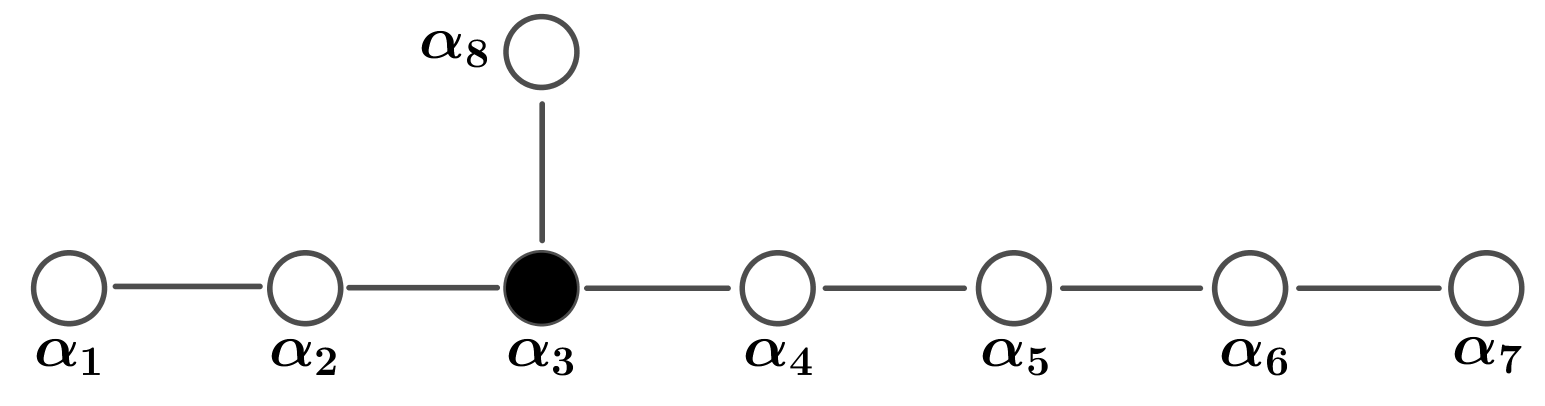}
  \end{center}
  \caption{$E_8$ Dynkin diagram}
   \label{E8dynkin}
  \end{figure}

In this section we conclude our analysis of simple flops by dealing with the highest length case, i.e.\ a flop of length 6 arising from a $E_8$ singularity deformed over the plane $\mathbb{C}_w$. We choose the Higgs $\Phi \in E_8$ in such a way to resolve only the central node of the $E_8$ Dynkin diagram as depicted in Figure~\ref{E8dynkin}.
\indent According to the principles outlined in previous sections, the Higgs field resolving the central node must lie in the Levi subalgebra defined by:
\begin{equation}\label{E8 length 6 levi}
\mathcal{L} = A_4^{(4,5,6,7)}\oplus A_2^{(1,2)}\oplus A_1^{(8)}\oplus \langle \alpha_3^*\rangle,
\end{equation}
where as usual the upper indices label the simple roots. Again we choose $\Phi$ of the following form:
\begin{equation}
\setlength{\jot}{16pt}
\renewcommand{\arraystretch}{1}
\begin{split}
& \Phi|_{{A_4}^{(4,5,6,7)}} = \left(\begin{array}{ccccc}
0 & 1 & 0 & 0 & 0 \\
0 & 0 & 1 & 0 & 0 \\
0 & 0 & 0 & 1 & 0 \\
0 & 0 & 0 & 0 & 1 \\
c_{4567}w & 0 & 0 & 0 & 0 \\ 
\end{array}\right) =e_{\alpha_4} +e_{\alpha_5} + e_{\alpha_6} +e_{\alpha_7} - c_{4567}w \left[\left[\left[e_{-\alpha_1},e_{-\alpha_2}\right],e_{-\alpha_3}\right],e_{-\alpha_4}\right], \\
& \Phi|_{{A_2}^{(1,2)}} = \left(\begin{array}{ccc}
0 & 1 & 0  \\
0 & 0 & 1 \\
c_{12}w & 0 & 0 \\
\end{array} \right)= e_{\alpha_1} +e_{\alpha_2} +  c_{12}w \left[e_{-\alpha_1},e_{-\alpha_2}\right],\\
& \Phi|_{{A_1}^{(8)}} = \left(\begin{array}{cc}
0 & 1 \\
c_8w & 0 \\
\end{array}\right)= e_{\alpha_8} + c_8w\hspace{0.1cm}e_{-\alpha_8}, \\
&\Phi\rvert_{\langle\alpha_3^*\rangle} = c_3w\, \langle\alpha_3^*\rangle.
\end{split} 
\end{equation}

To make the equation more readily understandable, we set the parameters to a specific value $c_3=0,c_8=1,c_{12}=-1,c_{4567}=1$. In this way we obtain the threefold
\begin{equation}
\scalemath{0.95}{
x^2+y^3+z^5+\frac{89 w^7}{144}+\frac{w^6}{864}+\frac{55 w^5 y}{12}-\frac{w^4 y}{48}-\frac{17 w^4 z^2}{24}-\frac{11 w^3 z^3}{12}-\frac{7}{2} w^2 y z^2-w y z^3=0\:,}
\end{equation}
where we neglected terms of high degree, irrelevant for the singularity at $w=0$.

\

{\bf Zero modes.} 
We perform the mode counting explicitly, independently for each irreducible representation arising from the adjoint $\boldsymbol{248}$ of $E_8$, branched under the Levi subalgebra \eqref{E8 length 6 levi}. The decomposition reads:
\begin{equation}
\begin{split}
\boldsymbol{248}\hspace{0.2cm} =\hspace{0.2cm} &(\boldsymbol{24},\boldsymbol{1},\boldsymbol{1})_{0}\oplus(\boldsymbol{1},\boldsymbol{8},\boldsymbol{1})_{0}\oplus(\boldsymbol{1},\boldsymbol{1},\boldsymbol{3})_{0}\oplus(\boldsymbol{1},\boldsymbol{1},\boldsymbol{1})_{0}\hspace{0.1cm}\oplus \\
& (\boldsymbol{5},\boldsymbol{\overline{3}},\boldsymbol{2})_{1}\oplus(\boldsymbol{\overline{5}},\boldsymbol{3},\boldsymbol{2})_{-1}\oplus(\boldsymbol{10},\boldsymbol{3},\boldsymbol{1})_{2}\oplus(\boldsymbol{\overline{10}},\boldsymbol{\overline{3}},\boldsymbol{1})_{-2}\hspace{0.1cm}\oplus \\
& (\boldsymbol{\overline{10}},\boldsymbol{1},\boldsymbol{2})_{3}\oplus(\boldsymbol{10},\boldsymbol{1},\boldsymbol{2})_{-3}\oplus(\boldsymbol{\overline{5}},\boldsymbol{\overline{3}},\boldsymbol{1})_{4}\oplus(\boldsymbol{5},\boldsymbol{3},\boldsymbol{1})_{-4}\hspace{0.1cm}\oplus \\
& (\boldsymbol{1},\boldsymbol{3},\boldsymbol{2})_{5}\oplus(\boldsymbol{1},\boldsymbol{\overline{3}},\boldsymbol{2})_{-5}\oplus(\boldsymbol{5},\boldsymbol{1},\boldsymbol{1})_{6}\oplus(\boldsymbol{\overline{5}},\boldsymbol{1},\boldsymbol{1})_{-6}
\end{split}
\end{equation}
Applying the Mathematica routine presented in Appendix~\ref{Appendix B}, we find the zero modes in Table~\ref{modes e86}.
\renewcommand{\arraystretch}{1.5}
\begin{table}[t]
  \centering
  \begin{tabular}{ |p{2.cm}||p{2.cm}|p{2.cm}||p{2.cm}|}
    \hline
    $R^{\mathcal L}$ & $\mathbb C[w]/(w)$ & $\mathbb C[w]/(w^2)$ &$\#_{\rm zero\,modes}$\\
    \hline
    $(\boldsymbol{24},\boldsymbol{1},\boldsymbol{1})_{0}$ & 0     &  0 &  0             \\
    $(\boldsymbol{1},\boldsymbol{8},\boldsymbol{1})_{0}$ & 0     &  0   &  0            \\
    $(\boldsymbol{1},\boldsymbol{1},\boldsymbol{3})_{0}$ & 0     &  0   &  0            \\
    $(\boldsymbol{1},\boldsymbol{1},\boldsymbol{1})_{0}$ & 0     &  0    &  0           \\
    \hline
    $(\boldsymbol{5},\boldsymbol{\overline{3}},\boldsymbol{2})_{1}$ & 6     &
                                                                           0  &  6  \\
    $(\boldsymbol{\overline{5}},\boldsymbol{3},\boldsymbol{2})_{-1}$ & 6     &  0    &  6 \\
    $(\boldsymbol{10},\boldsymbol{3},\boldsymbol{1})_{2}$ & 6 &0 &  6  \\
    $(\boldsymbol{\overline{10}},\boldsymbol{\overline{3}},\boldsymbol{1})_{-2}$ & 6     &  0     &  6 \\
    \hline                                                             
    $(\boldsymbol{\overline{10}},\boldsymbol{1},\boldsymbol{2})_{3}$ & 4     &  0  &  4\\
    $(\boldsymbol{10},\boldsymbol{1},\boldsymbol{2})_{-3}$ & 4
                                          &  0   &  4  \\
    $(\boldsymbol{\overline{5}},\boldsymbol{\overline{3}},\boldsymbol{1})_{4}$ & 3 &  0  &  3\\
    $(\boldsymbol{5},\boldsymbol{3},\boldsymbol{1})_{-4}$ & 3     &  0   &  3  \\
    \hline
    $(\boldsymbol{1},\boldsymbol{3},\boldsymbol{2})_{5}$ & 2     &  0  &  2\\
    $(\boldsymbol{1},\boldsymbol{\overline{3}},\boldsymbol{2})_{-5}$ & 2
                                          &  0  &  2   \\
    $(\boldsymbol{5},\boldsymbol{1},\boldsymbol{1})_{6}$ & 1     &  0   &  1\\
    $(\boldsymbol{\overline{5}},\boldsymbol{1},\boldsymbol{1})_{-6}$ & 1     &  0  &  1   \\
    \hline
  \end{tabular}.
  \caption{Five-dimensional modes for $E_8$ length six simple flop.}
  \label{modes e86}
\end{table}
We find a total of 44 localized modes:
  \begin{itemize}
  \item one hyper with charge six, inside $(\textbf{5},\textbf{1},\textbf{1})_{6} \oplus
    (\overline{\textbf{5}},\textbf{1},\textbf{1})_{-6}$;
  \item two hypers with charge five, inside $(\textbf{1},\textbf{3},\textbf{2})_{5} \oplus
    (\textbf{1},\overline{\textbf{3}},\textbf{2})_{-5}$;
  \item three hypers with charge four, inside $(\overline{\textbf{5}},\overline{\textbf{3}},\textbf{1})_{4} \oplus (\textbf{5},\textbf{3},\textbf{1})_{-4}$;
  \item four hypers with charge three inside $(\overline{\textbf{10}},\textbf{1},\textbf{2})_{3} \oplus (\textbf{10},\textbf{1},\textbf{2})_{-3}$;
  \item six hypers with charge two inside $(\textbf{10},\textbf{3},\textbf{1})_{2}\oplus
    (\overline{\textbf{10}},\overline{\textbf{3}},\textbf{1})_{-2}$;
  \item six hypers with charge one inside $(\textbf{5},\overline{\textbf{3}},\textbf{2})_{1}\oplus
    (\overline{\textbf{5}},\textbf{3},\textbf{2})_{-1}$.
  \end{itemize}
In terms of the GV invariants, one then reads  
\begin{equation}\label{GVFlop3-genL6}
n^{g=0}_1=6, \qquad n^{g=0}_2=6, \qquad n^{g=0}_3=4, \qquad n^{g=0}_4=3, \qquad n^{g=0}_5=2 \quad\mbox{ and }\quad n^{g=0}_6=1 \:.\nonumber
\end{equation}

\

Finally, analyzing the rank of the matrix $B_d$, we find that no enhancement in the number of localized modes at $w=0$ is feasible without generating a non-isolated singularity.

\section{Conclusions}\label{Sec:conclusions}
\indent In the present work, we constructed and studied the GV invariants of examples of threefold simple flops of any length. Building on our previous results based on the M-theory/Type IIA duality \cite{Collinucci:2021ofd,Collinucci:2021wty}, we introduced techniques to deal with intrinsically non-perturbative cases, where the Type IIA limit is not available, involving one-parameter families of deformed $E_6,E_7,E_8$ singularities. By means of a group-theoretic computational algorithm (included in an ancillary Mathematica file), the determination of the GV invariants for flops of any length has been reduced to a linear algebra problem. Our method is explicit, and also enables us to detect special cases of threefold simple flops exhibiting enhanced GV invariants.\\
\indent Furthermore, from a physical perspective, our work allows to study the dynamics of M-theory on threefold simple flops, which are notably non-toric, characterizing the Higgs Branches of the resulting 5d SCFTs as complex algebraic varieties. This is possible thanks to the correspondence between the GV invariants of threefold simple flops and BPS states arising from M2-branes wrapped on the curve inflated by the resolution of the simple flops. Indeed, M2-brane states descend to hypermultiplets in 5d, whose charges under the flavor group are determined by the degrees of the GV invariants, which we computed explicitly in flops of all lengths. In particular, we are able to fully characterize the action of the flavor symmetries on the Higgs Branch.\\
\indent These results pave the way for a range of engaging perspectives. For example, it would be interesting to apply our methods to the so-called $(A_j,E_k)$ (with $k=6,7,8$) singularities, which require an intrinsically non-perturbative description. Furthermore, achieving a more thorough understanding of the construction of general Higgs backgrounds would prove hugely profitable for the study of more general one-parameter fibrations of exceptional singularities.

\section*{Acknowledgments}

A.C.~is a Research Associate of the Fonds de la Recherche Scientifique F.N.R.S.~(Belgium). The work of A.C.~is partially supported by IISN - Belgium (convention 4.4503.15), and supported by the Fonds de la Recherche Scientifique - F.N.R.S.~under Grant CDR J.0181.18. 
A.S. and R.V. acknowledge support by INFN Iniziativa Specifica ST\&FI. M.D.M. acknowledges support by INFN Iniziativa Specifica GAST.

\appendix

\section{The threefold equation: The Katz-Morrison method}\label{App:Threefold}

Let us explain the method for obtaining a simultaneous resolution developed by Katz and Morrison \cite{Katz:1992aa}. 

One can start from the family over $\mathfrak{t}$ \eqref{deformed ADE singularities}, that is the expression of the  versal deformation of the ADE singularity in which the deformation parameters are functions of the $t_i$'s that encode the volumes of the 2-cycles.  To write down the equation of the fibration over $\mathfrak{t}/\mathcal{W}'$, one first needs to write the $\mathcal{W}'$ invariant coordinates $\varrho_i$'s in terms of the $t_i$'s; inverting these relations and plugging the result in \eqref{deformed ADE singularities} one obtains the hypersurface equation of the partial simultaneous resolution \cite{Katz:1992aa}. 

Let us immediately clarify this statement with the example of the deformed $A_3$ singularity we have introduced before. According to \eqref{deformed ADE singularities} the defining equation reads:
\begin{equation}\label{A3 ti}
x^2+y^2+(z-t_1)(z-t_2)(z-t_3)(z-t_4)=0 \quad\quad t_1+t_2+t_3+t_4=0.
\end{equation}
Now suppose again that we wish to produce a simple flop in which only the central node is resolved. The original Weyl group $\mathcal{W} = S_4$, acts as the permutation group on the four $t_i$'s. Here, the desired simultaneous resolution defines for us the subgroup $\mathcal{W}'=\mathbb{Z}_2\times \mathbb{Z}_2$, which acts by exchange of $t_1 \leftrightarrow t_2$ and $t_3 \leftrightarrow t_4$, respectively. It is easy to see that the $\mathcal{W}'$-invariant coordinates are:
\begin{equation}
 \tilde{s}_1= t_1\cdot t_2  \,, \qquad \tilde{s}_2=  t_1+t_2=-t_3-t_4  \,,\qquad \tilde{s}_3 = -t_1 t_3 - t_2t_3-t_3^2 = t_3\cdot t_4
\end{equation}
and comparing them to the expression \eqref{A3 ti} in terms of the $t_i$ it is a matter of mechanical computation\footnote{In practice, once one has the expression of the $\tilde{s}_i$ as functions of the $t_i$, one can write down the most general deformed $A_3$ singularity invariant under $\mathcal{W}'$ in terms of the $\varrho_i$ with unfixed coefficients. Comparing its terms degree by degree in the $t_i$ with expression \eqref{A3 ti} the correct coefficients can be easily extracted.} to extract the fibration over $\mathfrak{t}/\mathcal{W}'$:
\begin{equation}\label{A3familyExample}
x^2+y^2+(z^2-\tilde{s}_2z +\tilde{s}_1 )(z^2+\tilde{s}_2z +\tilde{s}_3 )=0.
\end{equation}
Finally, choosing a dependence $\tilde{s}_i(w)$ one gets a simple threefold flop. A well-known example is \emph{Reid's pagoda}, obtained by choosing:
\begin{equation}
\tilde{s}_2=0\,,  \quad  \tilde{s}_1 = -\tilde{s}_3 = w
\end{equation}
which gives the hypersurface $x^2+y^2+z^4-w^2$.

In principle, this procedure can be carried out for any deformed ADE singularity and any resolution pattern, although for the exceptional cases it can get particularly time-consuming.

\section{Casimirs for the equation of $E_r$ families}\label{Appendix A}
Let us consider the deformed $E_n$ singularities in equation \eqref{deformed ADE singularities}. In each one, we have $n$ deformation parameters:
\begin{equation}\label{En epsilon}
\begin{array}{c|cl}
\boldsymbol{E_6} & \epsilon_i & \text{for }i=2,5,6,8,9,12  \\
\boldsymbol{E_7} & \tilde{\epsilon}_i & \text{for }i=2,6,8,10,12,14,18\\
\boldsymbol{E_8} & \hat{\epsilon}_i & \text{for }i=2,8,12,14,18,20,24,30 \\
\end{array}
\end{equation}
Following Section~\ref{Sec:threefoldEq}, one computes their expression in terms of the Casimir $\chi_i^{E_r}(\Phi)$, with $\chi_i^{E_r}$ defined in \eqref{En casimirs}. 

Let us define (see \eqref{En casimirs})
\begin{equation}
c_{k_i}\equiv  \chi_i^{E_6}(\Phi),\qquad \tilde{c}_{k_i}\equiv  \chi_i^{E_7}(\Phi),\qquad \hat{c}_{k_i}\equiv  \chi_i^{E_8}(\Phi)\:.
\end{equation}

The result for the $E_6$ case is:
\begin{equation}\label{E6 epsilon}
\scalemath{1}{
\begin{split}
& \epsilon_2 = -\frac{c_{2}}{24}\\
& \epsilon_5 = \frac{c_{5}}{60}\\
& \epsilon_6 = \frac{c_{2}^3}{13824}-\frac{c_{6}}{144}\\
& \epsilon_8 = -\frac{c_{2}^4}{110592}+\frac{13 c_{2} c_{6}}{8640}-\frac{c_{8}}{240}\\
& \epsilon_9 = \frac{c_{9}}{756}-\frac{c_{2}^2 c_{5}}{11520}\\
& \epsilon_{12} =-\frac{c_{12}}{3240}+\frac{109 c_{2}^6}{4299816960}-\frac{847 c_{2}^3 c_{6}}{134369280}+\frac{109 c_{2}^2 c_{8}}{3732480}+\frac{13 c_{2} c_{5}^2}{466560}+\frac{61 c_{6}^2}{933120}.\\
\end{split}}
\end{equation}

For the $E_7$ case:
\begin{equation}\label{E7 epsilon}
\scalemath{1}{
\setlength{\jot}{16pt}
\begin{split}
\tilde{\epsilon}_2& =\frac{\tilde{c}_{2}}{18} \\
\tilde{\epsilon}_6 &=\frac{\tilde{c}_{2}^3}{139968}-\frac{\tilde{c}_{6}}{72} \\
\tilde{\epsilon}_8& =-\frac{7 \tilde{c}_{2}^4}{25194240}+\frac{11 \tilde{c}_{2} \tilde{c}_{6}}{16200}-\frac{\tilde{c}_{8}}{300} \\
\tilde{\epsilon}_{10} &=-\frac{2 \tilde{c}_{10}}{315}+\frac{\tilde{c}_{2}^5}{151165440}-\frac{17 \tilde{c}_{2}^2 \tilde{c}_{6}}{583200}+\frac{\tilde{c}_{2} \tilde{c}_{8}}{1400} \\
\tilde{\epsilon}_{12}& =-\frac{16 \tilde{c}_{10} \tilde{c}_{2}}{1148175}+\frac{\tilde{c}_{12}}{12150}-\frac{149 \tilde{c}_{2}^6}{10579162152960}+\frac{167 \tilde{c}_{2}^3 \tilde{c}_{6}}{3401222400}+\frac{737 \tilde{c}_{2}^2 \tilde{c}_{8}}{881798400}-\frac{31 \tilde{c}_{6}^2}{437400} \\
\tilde{\epsilon}_{14} &=\frac{8303 \tilde{c}_{10} \tilde{c}_{2}^2}{14935460400}-\frac{2201 \tilde{c}_{12} \tilde{c}_{2}}{217314900}+\frac{4 \tilde{c}_{14}}{62601}+\frac{11083 \tilde{c}_{2}^7}{24082404724998144}-\frac{11609 \tilde{c}_{2}^4 \tilde{c}_{6}}{5530387622400}\\
 &-\frac{1289 \tilde{c}_{2}^3 \tilde{c}_{8}}{1433804198400}+\frac{353 \tilde{c}_{2} \tilde{c}_{6}^2}{142242480}-\frac{31 \tilde{c}_{6} \tilde{c}_{8}}{1463400} \\
\tilde{\epsilon}_{18}& = \frac{12182634587 \tilde{c}_{10} \tilde{c}_{2}^4}{77806514663884339200}-\frac{564449 \tilde{c}_{10} \tilde{c}_{2} \tilde{c}_{6}}{3418744644000}+\frac{1844 \tilde{c}_{10} \tilde{c}_{8}}{3956880375}-\frac{27233975 \tilde{c}_{12} \tilde{c}_{2}^3}{11321053720935552}\\
 & +\frac{301 \tilde{c}_{12} \tilde{c}_{6}}{452214900}+\frac{307855 \tilde{c}_{14} \tilde{c}_{2}^2}{13588370378352}-\frac{2 \tilde{c}_{18}}{1507383}-\frac{886993691 \tilde{c}_{2}^9}{313644160640867419847393280}\\
 &+\frac{4713945967 \tilde{c}_{2}^6 \tilde{c}_{6}}{72026602145995788288000}-\frac{14715122551 \tilde{c}_{2}^5 \tilde{c}_{8}}{2334195439916530176000}-\frac{579011753 \tilde{c}_{2}^3 \tilde{c}_{6}^2}{23156700792822720000}\\
 &+\frac{2313866297 \tilde{c}_{2}^2 \tilde{c}_{6} \tilde{c}_{8}}{222355151645760000}-\frac{77393 \tilde{c}_{2} \tilde{c}_{8}^2}{3376537920000}-\frac{15011 \tilde{c}_{6}^3}{97678418400}.\\
\end{split}}
\end{equation}
For the $E_8$ case:
\begin{equation}\label{E8 epsilon}
\scalemath{0.8}{
\setlength{\jot}{13pt}
\begin{split}
 \hat{\epsilon}_2& =\frac{\hat{c}_{2}}{120} \\
 \hat{\epsilon}_8& =\frac{13 \hat{c}_{2}^4}{24883200000}-\frac{\hat{c}_{8}}{5760} \\
 \hat{\epsilon}_{12}& =\frac{\hat{c}_{12}}{181440}+\frac{101 \hat{c}_{2}^6}{3224862720000000}-\frac{\hat{c}_{2}^2 \hat{c}_{8}}{64512000} \\
 \hat{\epsilon}_{14}& =-\frac{71 \hat{c}_{12} \hat{c}_{2}}{798336000}+\frac{\hat{c}_{14}}{1108800}-\frac{2531 \hat{c}_{2}^7}{9029615616000000000}+\frac{103 \hat{c}_{2}^3 \hat{c}_{8}}{696729600000} \\
 \hat{\epsilon}_{18}& =-\frac{4451 \hat{c}_{12} \hat{c}_{2}^3}{689762304000000}+\frac{1523 \hat{c}_{14} \hat{c}_{2}^2}{12454041600000}-\frac{\hat{c}_{18}}{47174400}-\frac{26399 \hat{c}_{2}^9}{2080423437926400000000000}\\
 &+\frac{4747 \hat{c}_{2}^5 \hat{c}_{8}}{722369249280000000}+\frac{331 \hat{c}_{2} \hat{c}_{8}^2}{1672151040000} \\
 \hat{\epsilon}_{20}& = \frac{191071 \hat{c}_{12} \hat{c}_{2}^4}{2121019084800000000}+\frac{127 \hat{c}_{12} \hat{c}_{8}}{174569472000}-\frac{1165063 \hat{c}_{14} \hat{c}_{2}^3}{612738846720000000}+\frac{236627 \hat{c}_{18} \hat{c}_{2}}{434023349760000}\\
 &+\frac{10249681 \hat{c}_{2}^{10}}{61414099887587328000000000000}-\frac{2994007 \hat{c}_{2}^6 \hat{c}_{8}}{35540567064576000000000}-\frac{323371 \hat{c}_{2}^2 \hat{c}_{8}^2}{82269831168000000}-\frac{\hat{c}_{20}}{220809600}\\
 \hat{\epsilon}_{24}& =-\frac{193 \hat{c}_{12}^2}{17793312768000}+\frac{228270563 \hat{c}_{12} \hat{c}_{2}^6}{29320967828275200000000000}+\frac{234189517 \hat{c}_{12} \hat{c}_{2}^2 \hat{c}_{8}}{945465467240448000000}\\
 &-\frac{9171869023 \hat{c}_{14} \hat{c}_{2}^5}{52675933174824960000000000}-\frac{23281 \hat{c}_{14} \hat{c}_{2} \hat{c}_{8}}{9150846566400000}+\frac{561557071 \hat{c}_{18} \hat{c}_{2}^3}{8291582073815040000000}\\
 &+\frac{8268193432181 \hat{c}_{2}^{12}}{580761207304971815485440000000000000000}-\frac{20976434911 \hat{c}_{2}^8 \hat{c}_{8}}{3055351469407469568000000000000}\\
 &-\frac{16935675593 \hat{c}_{2}^4 \hat{c}_{8}^2}{33005339947302912000000000}-\frac{666323 \hat{c}_{2}^2 \hat{c}_{20}}{721337268326400000}+\frac{\hat{c}_{24}}{10061694720}-\frac{593 \hat{c}_{8}^3}{887354818560000} \\
  \hat{\epsilon}_{30}& = -\frac{636328729 \hat{c}_{12}^2 \hat{c}_{2}^3}{367646783551116410880000000}-\frac{189107437 \hat{c}_{12} \hat{c}_{14} \hat{c}_{2}^2}{277976001893990400000000}+\frac{2521 \hat{c}_{12} \hat{c}_{18}}{31907254579200000}\\
 &+\frac{122785779721089347 \hat{c}_{12} \hat{c}_{2}^9}{5354576379380206927872000000000000000000}+\frac{374760114643099 \hat{c}_{12} \hat{c}_{2}^5 \hat{c}_{8}}{685159914799807856640000000000000}\\
 &-\frac{199931513 \hat{c}_{12} \hat{c}_{2} \hat{c}_{8}^2}{94458563710156800000000}+\frac{28501673 \hat{c}_{14}^2 \hat{c}_{2}}{3860777804083200000000}-\frac{1634513578407571229 \hat{c}_{14} \hat{c}_{2}^8}{3206548401263100769075200000000000000000}\\
 &-\frac{3442332938170993 \hat{c}_{14} \hat{c}_{2}^4 \hat{c}_{8}}{593805259493166809088000000000000}+\frac{1223 \hat{c}_{14} \hat{c}_{8}^2}{112201334784000000}+\frac{15587535288859801 \hat{c}_{18} \hat{c}_{2}^6}{76346390506264304025600000000000000}\\
 &-\frac{1051350791 \hat{c}_{18} \hat{c}_{2}^2 \hat{c}_{8}}{1243310844834938880000000}+\frac{38736013334814563129113 \hat{c}_{2}^{15}}{919171413254131073937239231692800000000000000000000000}\\
 &-\frac{966205043352894287 \hat{c}_{2}^{11} \hat{c}_{8}}{46497194159854305977303040000000000000000000}-\frac{53516928494297557 \hat{c}_{2}^7 \hat{c}_{8}^2}{42002885419922588958720000000000000000}\\
 &-\frac{2159242595767 \hat{c}_{2}^5 \hat{c}_{20}}{737984035215212544000000000000}+\frac{21328481 \hat{c}_{2}^3 \hat{c}_{24}}{58332071437516800000000}+\frac{225239997090599 \hat{c}_{2}^3 \hat{c}_{8}^3}{119591548765057371340800000000000}\\
 &+\frac{72667 \hat{c}_{2} \hat{c}_{20} \hat{c}_{8}}{4518107320320000000}-\frac{\hat{c}_{30}}{1978376400000}.\\
\end{split}}
\end{equation}

\section{Milnor number and zero modes}\label{App Milnor}
From the point of view of the 5d theory coming from M-theory reduced on the singular space $X$,  the number of massless hypermultiplets gives the dimension $d_{HB}$ of the Higgs Branch (HB). This is due to the fact that for the threefolds we consider in this paper, there is no gauge symmetry (as the exceptional set has no codimension-1 loci).

The dynamics of M-theory on $X$ admits an alternative description in terms of the complex structure deformations of
  the threefold. 
When the singularity is deformed, a number $\mu$ of compact three-cycles blow up. Some of them intersect each other pairwise (these are called paired), while the other do not intersect any other compact three-cycle (these are called unpaired).   
   $\mu$ is called the Milnor number of the
  singularity $F(x,y,z,w)=0$ and it is the dimension, as complex vector space, of the
  Jacobian ring 
  \begin{equation}
    \mathcal J \equiv \frac{\mathbb C[x,y,w,z]}{(F, \partial_xF,
      \partial_yF, \partial_wF, \partial_zF)}.
  \end{equation}
  
  The dimension of the Higgs branch  is given by  counting appropriately the number of independent,
  dynamical complex structure 
  deformations that smooth the singularity. 
  For  non-compact Calabi-Yau varieties, the number $d_{HB}$ of independent, dynamical
  complex deformations coincides with \cite{Shapere:1999xr,Gukov:1999ya,Xie:2015rpa}
  \begin{equation}
    d_{HB} = \frac{\mu - \ell}{2} + \ell,
  \end{equation}
  where $\mu$ is the Milnor number and $\ell$ the number of ``unpaired''
  three-cycles.

  On general grounds, $\ell$
  coincides with the number of two-cycles inflated in the resolution, that in our case coincides with the dimension of $\mathcal{H}$ in \eqref{calH}. For the simple flop cases studied in this paper $\ell=1$ and hence 
$d^{\rm simple \, flop}_{HB} = \frac{\mu -1}{2} + 1$.

We have used this formula to check our results: we first compute the Milnor number for the examples in Section~\ref{Simple flop section} to compute the Higgs branch dimension $d^{\rm simple \, flop}_{HB}$; we then checked that this number coincided with the total number of hypermultiplets we computed.

\section{Mathematica code for computing the zero modes}\label{Appendix B}

In this section we will describe the ancillary Mathematica code, which can
be found at the \textit{arXiv} page of this paper. We loaded, together with
the paper, a zipped folder. The folder contains nine text files containing the positive and negative roots of the exceptional
algebras\footnote{For the $E_6$ algebra, the root vectors are in the $\textbf{27}$
  representation. For the $E_7,$ $E_8$ algebras the root vectors are in the adjoint
  representation.} and a basis of the Cartan subalgebras. We used, to
produce these matrices, the results of \cite{Deppisch,Cacciatori1, Cacciatori2}. These text files
should be saved in one of the folders of the variable \texttt{\$Paths} of
Mathematica. In the folder, one can find also the
notebook file "CodeHiggsBranchData.nb" which is divided into two sections. The first
section ``Main Code'' contains the routines that we used to produce the Higgs
branch data having, as input, the Higgs field describing the type IIA
dual of M-theory on the considered threefold. The second sections ``Examples'' contains the application of the functions defined in the section ``Main Code'' to the simple flops we analyzed in this paper. We now describe the main routines of the ancillary Mathematica code.

\paragraph{HbData function}

The main routine contained in
the code is
\begin{equation*}
  \label{eq:163}
  \footnotesize{\text{\texttt{HbData[ADE, rank, listhiggs, coeffhiggs, cartanhiggs, coeffcartan, listlevi]}}}.
\end{equation*}
The arguments of the function are
\begin{itemize}
\item \texttt{ADE}: is a \texttt{Symbol} to be picked among \texttt{"A, DD, E6, E7, E8"} and specifies
  the type of ADE algebra associated to the threefold.
\item \texttt{rank}: is a positive \texttt{Integer} that specifies the rank of the
  ADE algebra associated to the threefold.
\item \texttt{listhiggs}: is a \texttt{List} of \texttt{Lists}. Each sublist represents a root such
  that the Higgs field has a non-zero coefficient along the corresponding
  root-vector in $\mathfrak{g}$. The root vectors considered here do not lie
  in the Cartan subalgebra: the elements in the Cartan subalgebra will be separately input with the
  variables \texttt{cartanhiggs} and \texttt{coeffcartan}.
  The roots are described by their integer coefficients
  decomposition with respect to the simple roots. For example, keeping the
  notations in section  \ref{Length five flop section}, the root vector $v =
  \left[\left[e_{-\alpha_5},e_{-\alpha_6}\right]e_{-\alpha_7}\right]$
appearing in \eqref{Higgs tails length five} is expressed  as
\begin{equation*}
  \label{eq:73app}
\left\{0,0,0,0,-1,-1,-1,0\right\}.
\end{equation*}
To order the roots, we followed the labelling of Figures
\ref{E6dynkin}, \ref{E7dynkin}, 
  \ref{E8dynkinlength5} and \ref{E8dynkin} for the exceptional algebras. We labelled roots from the left to the
  right in the $A_r$ Dynkin diagram. For the $D_r$ case, the first $r-3$ integers are
  associated to the $A_{r-3}$ subalgebra associated to the longest tail of
  the $D_r$ Dynkin diagram (again, labelling roots from left to right),
  the  third last integer number appearing  in the sublists of \texttt{listhiggs}
  is associated to the
  trivalent root,  the last two integers
  are associated to
  the two $A_1$ short tails of the $D_r$ diagram.

  For example, if we again consider the Higgs field of section  \ref{Length five flop section}, we have to insert, as input
  \begin{eqnarray*}
    \label{eq:172}
    \scriptstyle \text{\texttt{listhiggs}}&\scriptstyle = &
     \Big\{
    \text{ }\scriptstyle \red{\left\{0,0,0,0,1,0,0,0\right\},
    \left\{0,0,0,0,0,1,0,0\right\},
                                                \left\{0,0,0,0,0,0,1,0\right\}},
                                                 \nonumber \\
    &&  \quad \scriptstyle 
      \red{\left\{0,0,0,0,-1,-1,-1,0\right\}},
      \blue{ \left\{1,0,0,0,0,0,0,0\right\},
       \left\{0,1,0,0,0,0,0,0\right\}},
        \nonumber \\
    &&  \quad \scriptstyle 
    \blue{\left\{0,0,1,0,0,0,0,0\right\},
    \left\{0,0,0,0,0,0,0,1\right\},
    \left\{-1,-1,-1,0,0,0,0,-1\right\}}
    \Big \}, \nonumber \\
  \end{eqnarray*}
  where the red elements correspond to $\Phi\rvert_{A_3^{(5,6,7)}}$, and
  the blue elements to $\Phi\rvert_{A_4^{(1,2,3,8)}}$ in \eqref{Higgs tails length five}.
  
  The root system, in our convention, can be printed on screen calling the function \texttt{PrintRootSystem[ADE,rank]} (the first argument being again the ADE type of $\mathfrak{g}$, and the second argument its rank).
\item \texttt{coeffhiggs}: is a \texttt{List} containing the coefficients
  corresponding to the elements of \texttt{listhiggs}. If we again consider
  the Higgs field of \eqref{Higgs tails length five} we have 
  \begin{equation*}
    \label{eq:174}
    \text{\texttt{coeffhiggs}} = \left\{\red{1, 1, 1,  w
      c_{567}}, \blue{1, 1, 1, 1, w c_{1238}}\right\}.
  \end{equation*}
\item \texttt{cartanhiggs}: is a \texttt{List} of positive \texttt{Integers} $n_i$, with
  $n_i$ = 1, ..., \texttt{rank}, describing the elements of the Cartan
  subalgebra $\mathcal C$ of $\mathfrak{g}$ along which the Higgs field has a
  non-zero coefficient. The generators of the Cartan subalgebra has been chosen
  as the dual elements $\alpha_j^*$ of the simple roots.  For example,
  let's consider again the $E_8$ algebra. We know by construction that the
  Higgs field has to preserve the Cartan element dual  to $\alpha_4$ in Figure \ref{E8dynkinlength5}. This Cartan element is defined as the one that  commutes with all
the simple root vectors different from $e_{\alpha_4}$, and
\begin{equation*}
  \label{eq:181}
  \left[\alpha_4^*,e_{\alpha_4}\right] = 1.
\end{equation*}
In order to pick a Higgs field with a non-zero component along $\alpha_4$
we then input
  \begin{equation*}
    \label{eq:178}
    \text{\texttt{cartanhiggs}}= \left\{4\right\}.
\end{equation*}

For non-simple flops, where we have more than one $\mathbb P^1$ resolved,
the input
\begin{equation*}
      \text{\texttt{cartanhiggs}}=
  \left\{n_1,...,n_f\right\},
\end{equation*}
corresponds to turning on a Higgs v.e.v. along the roots $\alpha_{n_1}^*,...,\alpha_{n_f}^*$.
\item \texttt{coeffcartan}: is a \texttt{List} of \texttt{Symbols} describing the
  coefficients corresponding to the elements of \texttt{cartanhiggs}. In the
  previous example, if we input
  \begin{equation*}
     \text{\texttt{cartanhiggs}}=
   \left\{4\right\}, \quad  \text{\texttt{coeffcartan}}=
  \left\{w c_4\right\},
\end{equation*}
we picked the Higgs to have a coefficient $w c_4$ along $\alpha_4^*$.
\item \texttt{listlevi}: is a \texttt{List} of \texttt{Integers} describing the simple roots
  that labels the Levi subalgebra. In the previous example, we picked the
  root vectors all
  residing in the Levi subalgebra labeled by all the roots of $E_8$ but the resolved
  one. The fact that the Levi subalgebra associated to the Higgs field
  \eqref{eq:164} is $L = A_{3}^{(5,6,7)} \oplus
  A_4^{(1,2,3,8)} \oplus \langle \alpha_4^* \rangle$ is explained in
  details in section \ref{Length five flop section} but can be also read off from our choice of the variable
  \texttt{listhiggs}. Indeed, the simple roots labeling the Levi subalgebra
  (in this case, $\left\{1,2,3,5,6,7,8\right\}$) are the minimum amount of
  simple roots we need to generate (with integer coefficients) all the root
  we input in \texttt{listhiggs}. All the Levi subalgebras contains \cite{Collingwood}
  the Cartan subalgebra, hence the data \texttt{cartanhiggs} and
  \texttt{coeffcartan} do not modify the Levi subalgebra datum.
\end{itemize}

The function has a void output and prints all the data we need to determine the action of the
spontaneously broken flavor symmetry on the Higgs Branch (or, in other
words, the Gopakumar-Vafa invariants of the small resolution of the
singularity, labeled by their degrees).

As an example we
report here a part of the output of the length-five simple flop case we
analyzed in section \ref{Length five flop section}. The whole output is contained in the
\texttt{Examples} section of the ancillary Mathematica file. In this case,
the output contains many blocks (one for each irreducible representation of the branching of
$\mathfrak{g}$ with respect to the Levi subalgebra \eqref{eq:164}) of the
following type:
\begin{center}
\includegraphics[scale=1.]{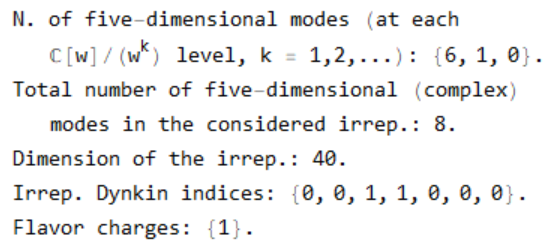}
\end{center}
The first three lines report the number of five-dimensional modes localized in
the considered irreducible representation. In this case, we read the list
$\left\{6,1,0\right\}$, this means that we have six modes localized in
$\mathbb C[w]/(w)$, one mode localized in $\mathbb C[w]/(w^2)$ and zero in
$\mathbb C[w]/(w^k)$ with $k > 2$. The
overall number of complex-valued modes is, hence, $6 * 1 + 1 * 2 = 8$ (indeed, a
mode inside $\mathbb C[w]/(w^k)$ counts for $k$ complex-valued
five-dimensional degrees of freedom). The last three lines tell us:
\begin{itemize}
\item  \textit{The complex dimension of the considered irreducible representation.} In the example, we have $40$.
\item \textit{The Dynkin indices of the highest weight state of the representation.} In the example, we read
  \begin{equation*}
    \left\{0,0,\red{1},\red{1},0,0,0\right\}.
  \end{equation*}
  Since the variable \texttt{listlevi} is $\left\{1,2,\red{3},\red{5},6,7,8\right\}$,
  the output is telling us that the lowest weight has weight $1$ with
  respect to the third root of Figure \ref{E8dynkinlength5}, $1$ with respect to the fourth root and
  zero with respect to all the other roots labelling the Levi subalgebra.
\item \textit{The charges of these modes with respect to the flavor group
  generators.} The generators of the flavor group are the Catan elements $\alpha_i^*$  that are dual to the roots
  that gets resolved. In this case, we see that all the five-dimensional
  modes in the considered irreducible representation have weight one with
  respect to the $\mathfrak{u}(1)$ flavor symmetry associated to the
  resolved node of Figure \ref{E8dynkinlength5}. This implies, in terms of
  the Higgs Branch geometry, that there is a spontaneously
  broken $\mathfrak{u}(1)$ symmetry, that acts with charge one on the modes
  localized in the considered irreducible representation.
\end{itemize}

\paragraph{ExtractRootDec function}

The second main routine is
\begin{equation*}
\text{\texttt{ExtractRootDec[ADE, rank, higgs]}}.  
\end{equation*}
The routine can be used to analyze, in the language of this paper, the
Higgs fields we presented in \cite{Collinucci:2021ofd,DeMarco:2021try}.
The first two arguments of the function are again, as in the \texttt{HbData}
function, the ADE type of $\mathfrak{g}$ and its rank. The third argument is
a matrix, representing the Higgs field. The Higgs field has to be input
\begin{itemize}
\item in the fundamental representations for the $A_r$, $D_r$ cases
  (following the notations in \cite{Collingwood});
\item in the $\textbf{27}$ representation for the $E_6$ case;
\item in the adjoint representation for the $E_7$, $E_8$ cases.
\end{itemize}

The output is a List containing 
 the arguments of the function \texttt{HbData} (ordered as in the function call process). In other words, the
first argument of the output List will be \texttt{ADE}, the second will be
\texttt{rank}, the third \texttt{listhiggs} and so on. The function does
not output the datum \texttt{listlevi}, which has to be added manually by
the user when \texttt{HbData} is called. We remark again that the simple roots labelling the Levi subalgebra
  are the minimum amount of
  simple roots we need to generate (with integer coefficients) all the roots
  in \texttt{listhiggs} (namely, appearing in the third element of the output of \texttt{ExtractRootDec[ADE, rank, higgs]}).



\begin{thebibliography}{10}

\bibitem{Katz:1992aa}
S.~Katz and D.~R. Morrison, ``Gorenstein threefold singularities with small
  resolutions via invariant theory for weyl groups,'' {\em J. Alg. Geom.} {\bf
  1} (1992) 449--530, \href{http://arXiv.org/abs/alg-geom/9202002}{{\tt
  alg-geom/9202002}}.

\bibitem{Collinucci:2021wty}
A.~Collinucci, A.~Sangiovanni, and R.~Valandro, ``{Genus zero Gopakumar-Vafa
  invariants from open strings},'' {\em JHEP} {\bf 09} (2021) 059,
  \href{http://arXiv.org/abs/2104.14493}{{\tt 2104.14493}}.

\bibitem{Collinucci:2021ofd}
A.~Collinucci, M.~De~Marco, A.~Sangiovanni, and R.~Valandro, ``{Higgs branches
  of 5d rank-zero theories from geometry},'' {\em JHEP} {\bf 10} (2021),
  no.~18, 018, \href{http://arXiv.org/abs/2105.12177}{{\tt 2105.12177}}.

\bibitem{Curto:aa}
C.~Curto and D.~Morrison, ``Threefold flops via matrix factorization,'' {\em
  Journal of Algebraic Geometry} {\bf 22} (12, 2006).

\bibitem{Karmazyn:2017aa}
J.~Karmazyn, ``The length classification of threefold flops via noncommutative
  algebras,'' {\em Advances in Mathematics} {\bf 343} (2019) 393--447.

\bibitem{Cachazo:2001gh}
F.~Cachazo, S.~Katz, and C.~Vafa, ``Geometric transitions and n=1 quiver
  theories,'' \href{http://arXiv.org/abs/hep-th/0108120}{{\tt hep-th/0108120}}.

\bibitem{Collinucci:2018aho}
A.~Collinucci, M.~Fazzi, and R.~Valandro, ``{Geometric engineering on flops of
  length two},'' {\em JHEP} {\bf 04} (2018) 090,
  \href{http://arXiv.org/abs/1802.00813}{{\tt 1802.00813}}.

\bibitem{Collinucci:2019fnh}
A.~Collinucci, M.~Fazzi, D.~R. Morrison, and R.~Valandro, ``{High electric
  charges in M-theory from quiver varieties},'' {\em JHEP} {\bf 11} (2019) 111,
  \href{http://arXiv.org/abs/1906.02202}{{\tt 1906.02202}}.

\bibitem{Cecotti:2010bp}
S.~Cecotti, C.~Cordova, J.~J. Heckman, and C.~Vafa, ``T-branes and monodromy,''
  {\em JHEP} {\bf 1107} (2011) 030, \href{http://arXiv.org/abs/1010.5780}{{\tt
  1010.5780}}.

\bibitem{Dixmier}
J.~Dixmier, ``Enveloping algebras,'' {\em North-Holland, Amsterdam} (1977).

\bibitem{Okubo}
S.~Okubo and J.~Patera, ``General indices of representations and casimir
  invariants,'' {\em J. Math. Phys. 25} {\bf 25} (1984).

\bibitem{Kollar}
H.~Clemens, J.~Koll{\'a}r, and S.~Mori, ``Higher-dimensional complex
  geometry,'' {\em Ast{\'e}risque} (1989).

\bibitem{Gopakumar:1998ii}
R.~Gopakumar and C.~Vafa, ``{M theory and topological strings. 1.},''
  \href{http://arXiv.org/abs/hep-th/9809187}{{\tt hep-th/9809187}}.

\bibitem{Gopakumar:1998ki}
R.~Gopakumar and C.~Vafa, ``{On the gauge theory / geometry correspondence},''
  {\em Adv. Theor. Math. Phys.} {\bf 3} (1999) 1415--1443,
  \href{http://arXiv.org/abs/hep-th/9811131}{{\tt hep-th/9811131}}.

\bibitem{Gopakumar:1998jq}
R.~Gopakumar and C.~Vafa, ``{M theory and topological strings. 2.},''
  \href{http://arXiv.org/abs/hep-th/9812127}{{\tt hep-th/9812127}}.

\bibitem{BrownWemyss}
G.~Brown and M.~Wemyss, ``{Gopakumar-Vafa invariants do not determine flops},''
  {\em Commun. Math. Phys.} {\bf 361} (2018), no.~1, 143--154,
  \href{http://arXiv.org/abs/1707.01150}{{\tt 1707.01150}}.

\bibitem{Collingwood}
D.~H. Collingwood and W.~M. McGovern, ``Nilpotent orbits in semisimple lie
  algebra: An introduction,'' {\em Chapman and Hall/CRC} (1993).

\bibitem{Shapere:1999xr}
A.~D. Shapere and C.~Vafa, ``{BPS structure of Argyres-Douglas superconformal
  theories},'' \href{http://arXiv.org/abs/hep-th/9910182}{{\tt
  hep-th/9910182}}.

\bibitem{Gukov:1999ya}
S.~Gukov, C.~Vafa, and E.~Witten, ``{CFT's from Calabi-Yau four folds},'' {\em
  Nucl. Phys. B} {\bf 584} (2000) 69--108,
  \href{http://arXiv.org/abs/hep-th/9906070}{{\tt hep-th/9906070}}. [Erratum:
  Nucl.Phys.B 608, 477--478 (2001)].

\bibitem{Xie:2015rpa}
D.~Xie and S.-T. Yau, ``{4d N=2 SCFT and singularity theory Part I:
  Classification},'' \href{http://arXiv.org/abs/1510.01324}{{\tt 1510.01324}}.

\bibitem{Deppisch}
T.~Deppisch, ``{E6Tensors: A Mathematica Package for E6 Tensors},'' {\em
  Comput. Phys. Commun.} {\bf 213} (2017) 130--135,
  \href{http://arXiv.org/abs/1605.05920}{{\tt 1605.05920}}.

\bibitem{Cacciatori1}
S.~L. Cacciatori, F.~Dalla~Piazza, and A.~Scotti, ``{$E_7$ groups from
  octonionic magic square},'' {\em Adv. Theor. Math. Phys.} {\bf 15} (2011),
  no.~6, 1605--1654, \href{http://arXiv.org/abs/1007.4758}{{\tt 1007.4758}}.

\bibitem{Cacciatori2}
S.~L. Cacciatori, F.~D. Piazza, and A.~Scotti, ``{A simple $E_8$
  construction},'' \href{http://arXiv.org/abs/1207.3623}{{\tt 1207.3623}}.

\bibitem{DeMarco:2021try}
M.~De~Marco and A.~Sangiovanni, ``{Higgs Branches of rank-0 5d theories from
  M-theory on (A$_{j}$, A$_{l}$) and (A$_{k}$, D$_{n}$) singularities},'' {\em
  JHEP} {\bf 03} (2022) 099, \href{http://arXiv.org/abs/2111.05875}{{\tt
  2111.05875}}.

\end{thebibliography}

\providecommand{\href}[2]{#2}

\end{document}